\documentclass[a4paper,fleqn,usenatbib]{mnras}

\usepackage{mathptmx}

\usepackage[T1]{fontenc}
\usepackage{ae,aecompl}

\usepackage{graphicx}	
\usepackage{amsmath}	
\usepackage{amssymb}	

\renewcommand\ion[2]{#1\uppercase\expandafter{\romannumeral#2}\relax}%
\def\eps@scaling{1.0}%
\newcommand\epsscale[1]{\gdef\eps@scaling{#1}}%
\newcommand\plotone[1]{%
 \typeout{Plotone included the file #1}
 \centering
 \leavevmode
 \includegraphics[width={\eps@scaling\columnwidth}]{#1}%
}%
\newcommand\plottwo[2]{{%
 \typeout{Plottwo included the files #1 #2}
 \centering
 \leavevmode
 \columnwidth=.45\columnwidth
 \includegraphics[width={\eps@scaling\columnwidth}]{#1}%
 \hfil
 \includegraphics[width={\eps@scaling\columnwidth}]{#2}%
}}%

\title[LWA1 Sky Survey]{The LWA1 Low Frequency Sky Survey}

\author[Dowell et al.]{
Jayce Dowell,$^{1}$\thanks{E-mail: jdowell@unm.edu}
Gregory B. Taylor,$^{1,2}$
Frank K. Schinzel,$^{1}$
Namir E. Kassim,$^{3}$ 
\newauthor
and Kevin Stovall$^{1}$
\\
$^{1}$Department of Physics and Astronomy, University of New Mexico, Albuquerque, NM  87131, USA\\
$^{2}$Greg Taylor is also an Adjunct Astronomer at the National Radio Astronomy Observatory.\\
$^{3}$Radio Astrophysics and Sensing Section, Naval Research Laboratory, Washington, D.C., 20375, USA
}

\date{Accepted XXX. Received YYY; in original form ZZZ}

\pubyear{2016}

\begin{document}
\label{firstpage}
\pagerange{\pageref{firstpage}--\pageref{lastpage}}
\maketitle

\begin{abstract}
We present a survey of the radio sky accessible from the first station of the Long Wavelength Array (LWA1).  Images are presented at nine frequencies between 35 and 80 MHz with spatial resolutions ranging from 4.7$^\circ$ to 2.0$^\circ$, respectively.  The maps cover the sky north of a declination of $-$40$^\circ$ and represent the most modern systematic survey of the diffuse Galactic emission within this frequency range.  We also combine our survey with other low frequency sky maps to create an updated model of the low frequency sky.  Due to the low frequencies probed by our survey, the updated model
better accounts for the effects of free-free absorption from Galactic
ionized Hydrogen.  A longer term motivation behind this survey is to understand the foreground emission that obscures the redshifted 21 cm transition of neutral hydrogen from the cosmic dark ages ($z>$10) and, at higher frequencies, the epoch of reionisation ($z>$6).
\end{abstract}

\begin{keywords}
Galaxy: general -- radio continuum: general --- techniques: interferometric
\end{keywords}



\section{Introduction}

In recent years much effort, both theoretical and observational, has
been invested in studies of the epoch of reionisation (EoR) and
the cosmic dark ages.  These two epochs represent the next frontier
for understanding the history of the Universe and, in particular, the
nature of the first structures to assemble.  Observations of the 21-cm
spin flip temperature from the cosmic dark ages will provide
information about the nature of the first stars and how they began to
ionise the intergalactic medium \citep{madau97,fur06,21Cosmology}.
This signal is expected to appear in the frequency range of 10 to 80
MHz with a strength of $\sim$100 mK.  The epoch of reionisation also
offers important constraints on the early universe through the time
evolution of the ionisation fraction of the intergalactic medium over
the frequency range from 100 to 200 MHz.

There are a variety of projects attempting to detect the dark ages
signal including the Large aperture Experiment to detect the Dark Ages
\citep[LEDA;][]{LEDA,FL}, Observing Cosmic Dawn with the Long
Wavelength Array \citep{FL,CosmicDawn}, LOfar COsmic-dawn Search
\citep[LOCOS;][]{LOCOS}, and the space-based Dark Ages Radio Explorer
\citep[DARE;][]{DARE}, among others.  Arrays such as the Precision
Array for Probing the Epoch of Reionisation \citep[PAPER;][]{PAPER}
and the Low Frequency Array \citep[LOFAR;][]{LOFAREOR} are attempting
to detect the EoR signal in the 100 to 200 MHz range and face similar
challenges to the dark ages studies due to contamination from bright 
Galactic and Extragalactic
foreground emission.  Since the foreground
emission varies from 200 K to 20,000 K in the pertinent
frequency range, all attempts to detect the
dark ages or EoR signals need a robust method of subtracting the
foregrounds.  Although there have been novel techniques suggested for
mitigating the foregrounds, i.e., delay spectrum filtering
\citep{PAPER2,ved12}, and generalized morphological component
analysis \citep{chap13}, information about the foregrounds is still
necessary in order to 
understand their residual contamination and to hone foreground
mitigation techniques.

The leading model for the sky in the frequency range of 20 to 200 MHz
is the Global Sky Model by \citet[GSM;][]{GSM}.  This model is based
upon a principal component analysis of 11 sky maps ranging in
frequency from 10 MHz to 94 GHz.  Of these 11 maps seven are above 1
GHz.  Below this frequency there are only four maps included in the
model: 10 MHz \citep{GSM10}, 22 MHz \citep{GSM22}, 45 MHz
\citep{GSM45A,GSM45M,Guzman11}, and 408 MHz \citep{GSM408}.  Thus, in
the GSM, the region of interest to both cosmic dawn and the epoch of
reionisation is largely extrapolated from available map data.
Furthermore, many of the best lower frequency surveys, while carefully
conducted, relied on analog techniques which can be significantly
improved upon utilising modern digital technology.

Although comparisons between the realisation of the GSM at 10, 22, and 45 MHz and the respective input maps show agreement at the 10\% level when averaged over large fractions of the sky, the disagreement in the spatial structure is larger and reaches $\sim$40\% in the case of 45 MHz.  There are also uncertainties in the accuracy of the input maps used and the accuracy of the model outside of these particular frequencies.  In addition, these three input maps offer limited sky coverage ranging from only 35\% to 96\% of the sky, and the GSM is limited to extrapolations from the higher frequencies in the unobserved regions.  The higher frequency maps cannot, by themselves, account for thermal absorption from the ionised ISM \citep{Kassim89} necessarily rendering extrapolations to lower frequencies inaccurate.  Indeed, the greatest discrepancies between the 10, 22, and 45 MHz input maps and the corresponding GSM realisations are found in and around the Galactic plane.

In order to overcome the limitations of the GSM and provide a set of
self-consistent maps at the frequencies of interest to the various
high redshift cosmological studies, we have undertaken a survey of the
sky between 35 and 80 MHz using the first station of the Long
Wavelength Array, LWA1.  The LWA1 is an ideal telescope for conducting
this survey due to a combination of sensitivity, observatory latitude,
and degree of aperture filling.  The resulting maps will provide a
direct comparison for the LEDA and Cosmic Dawn projects that are using
the LWA1 for data collection and can also constrain spectral continuum
models of the foreground emission used for higher frequency epoch of
reionisation studies. Although current studies of the EoR are trying
to detect angular fluctuations on sub-degree scales (smaller than the
resolution of the images presented here), these images will still be
of use in understanding the detailed spectral structure of foreground
emission.

Beyond the cosmological studies, accurate all-sky maps at these
frequencies are of interest for a wide variety of science cases.  For
radio astronomy, the sensitivity of LWA1 to diffuse synchrotron
emission and thermal \ion{H}{2} absorption relates directly to cosmic
ray physics \citep{Longair90,Webber90,Nord04}. In particular, the same
cosmic ray electrons generating the low frequency synchrotron emission
also produce diffuse gamma ray emission through \ion{H}{1} collisions,
as previously observed by the Fermi telescope
\citep{Ackermann12}. Models unifying our understanding of the origin
and acceleration of Galactic cosmic rays must be reconciled with the
spectral structure revealed in the LWA1 maps. At lower levels,
emission in the LWA1 maps relates to the unresolved extragalactic
background at both radio and higher energies. On more discrete scales,
the LWA1 maps can offer additional constraints for underlying features
such as the Fermi bubbles \citep{FermiB}.  Beyond radio astronomy, the
LWA1 maps can be brought to bear on geophysical measurements including
ionospheric remote sensing via imaging riometery \citep{det90}.

This paper is organised as follows: Section \ref{sec:obs} describes the LWA1, the data acquired for the survey, and the details of the data reduction techniques used.  Section \ref{sec:res} presents the all-sky maps in Galactic coordinates at nine frequencies between 35 and 80 MHz. Section \ref{sec:an} presents the comparisons with other maps and models as well as examines the spectral index of the sky.  Section \ref{sec:lfsm} presents our model of the low frequency radio sky.  We conclude with the prospects of extending the survey to both lower and higher frequencies in Section \ref{sec:conc}.

\section{Observations \& Data Analysis}
\label{sec:obs}

The data used for this survey was obtained with the first station of the
Long Wavelength Array \citep[LWA1,][]{LWA,FL}.  LWA1 is co-located
with the Karl G. Jansky Very Large Array in central New Mexico, USA
and operates over the 10 to 88 MHz frequency range.  The station
consists of 256 dual-polarisation dipole antennas spread over a 110 m
by 100 m north-south ellipse.  In addition to these 256 core dipoles there
are also five dual-polarisation dipole ``outriggers" located at
distances of $\sim$200 to 500 m from the core of the array.  The
signals from the dipoles are amplified and filtered by the analog
receivers (ARX) and then digitised and processed through two
subsystems: the beam former and the transient buffer.  The beam former
implements delay-and-sum beam forming and creates four independently
steerable dual-polarisation beams, each with two spectral tunings and
a bandwidth of up to 19.6 Msamples/s per tuning.  The transient buffer
provides a stream of raw voltage data from each antenna that is
suitable for cross-correlation in two modes: narrowband and wideband.
The narrowband mode provides 100 ksamples/s of complex voltages for
each antenna continuously, while the wideband mode provides up to 61
ms of the raw 12-bit digitiser output for every stand (dual
polarisation dipole antenna) over the entire observable sky with a
0.03\% duty cycle.

To cover a wide frequency range we have chosen to use the output of
the wideband transient buffer (TBW).  Although the 61-ms exposure
duration is short, the increased bandwidth allows the entire
wavelength range covered by the LWA1 to be recorded at once.  We
captured TBW data of the full sky visible at LWA1 once every 15
minutes for two full days: 2013 March 28 and 2013 April 30.  This
cadence over the two days allows for better sidereal time sampling of
the sky than a single day of capture and allows for the removal of the
Sun and radio frequency interference (RFI).  Six of the captures in
the 2013 April 30 data suffered from excessive RFI and consequentially
the corresponding sidereal time range was re-observed on 2013 May 11.
In total, our data consists of 193 captures from the 240 dipoles that
were fully operational at the time.  This provides a total integration
time of $\approx$11.8 s and a data volume of roughly 2.5 TB.  One
potential concern with this imaging approach is that the integration
time of 61 ms is too short to reach an interesting noise limit.
However, at 74 MHz the sky has a brightness temperature of $\sim$2,000
K and the LWA1 beam is $\sim$2$^\circ$.  Despite the short integration
time, the LWA1 images are expected to be confusion limited (see Table
\ref{tab:confusion}).  Since both the temperature of the sky and the
beam size scale roughly as $\nu^{-2}$, the surface brightness sensitivity is relatively
constant across all frequencies that are part of this survey.

\begin{table*}
	\centering
	\caption{Survey Parameters}
	\label{tab:confusion}
	\begin{tabular}{lccccc}
		\hline
		Frequency & Centre Frequency & Bandwidth & Beam Size$^a$ & Confusion & Thermal \\
		(MHz) & (MHz) & (kHz) & ($^\circ$) & Noise$^b$ (Jy beam$^{-1}$) & Noise$^c$ (Jy beam$^{-1}$) \\
		\hline
		35~MHz & 34.979 &  957 & 4.8 $\times$ 4.5 & 163 & 38 \\
		38~MHz & 38.042 &  957 & 4.5 $\times$ 4.1& 130 & 31 \\
                 40~MHz & 40.052 &  957 & 4.3 $\times$ 3.9 & 114 & 27 \\
                 45~MHz & 44.933 &  957 & 3.8 $\times$ 3.5 & 82  & 20 \\
                 50~MHz & 50.005 &  957 & 3.4 $\times$ 3.1 & 62  & 16 \\
                 60~MHz & 59.985 &  957 & 2.8 $\times$ 2.6 & 38  & 10 \\
                 70~MHz & 70.007 &  957 & 2.4 $\times$ 2.2 & 25  & 7 \\
                 74~MHz & 73.931 &  957 & 2.3 $\times$ 2.1 & 22  & 6 \\
                 80~MHz & 79.960 &  957 & 2.1 $\times$ 2.0 & 18  & 5 \\
		\hline
		\multicolumn{6}{p{.8\textwidth}}{$^a$  Assuming natural weighting.}\\
		\multicolumn{6}{p{.8\textwidth}}{$^b$  Calculated from source counts at 1.49 GHz \citep{confusion1,confusion2} and scaled to these frequencies using a spectral index of --0.7 \citep{confusion3}.}\\
		\multicolumn{6}{p{.8\textwidth}}{$^c$  Calculated assuming a 61 ms integration time and a sky temperature of 2000 K at 74 MHz with a spectral index of $-$2.5.} \\
	\end{tabular}
\end{table*}

Since the TBW data consist of time-domain voltages, they need to be cross-correlated to form visibilities that can be used for imaging.  The cross correlation was performed with the FX software correlator available as part of the LWA Software Library \citep[LSL;][]{LSL}.  The correlator was run on each capture with 1,024 channels using the cable and system delays stored in the LWA Data Archive\footnote{\url{http://lda10g.alliance.unm.edu/metadata/lwa1/ssmif/}}.  The correlation process also accounts for the frequency-dependent cable loss by correcting the visibility amplitudes.  The cables are buried to a minimum depth of 0.45 m and have been verified not to show significant effects of diurnal variation.  Each snapshot of the sky is phased to zenith and contains 28,920 baselines at a spectral resolution of 95.6 kHz.  

\begin{figure}
	\plotone{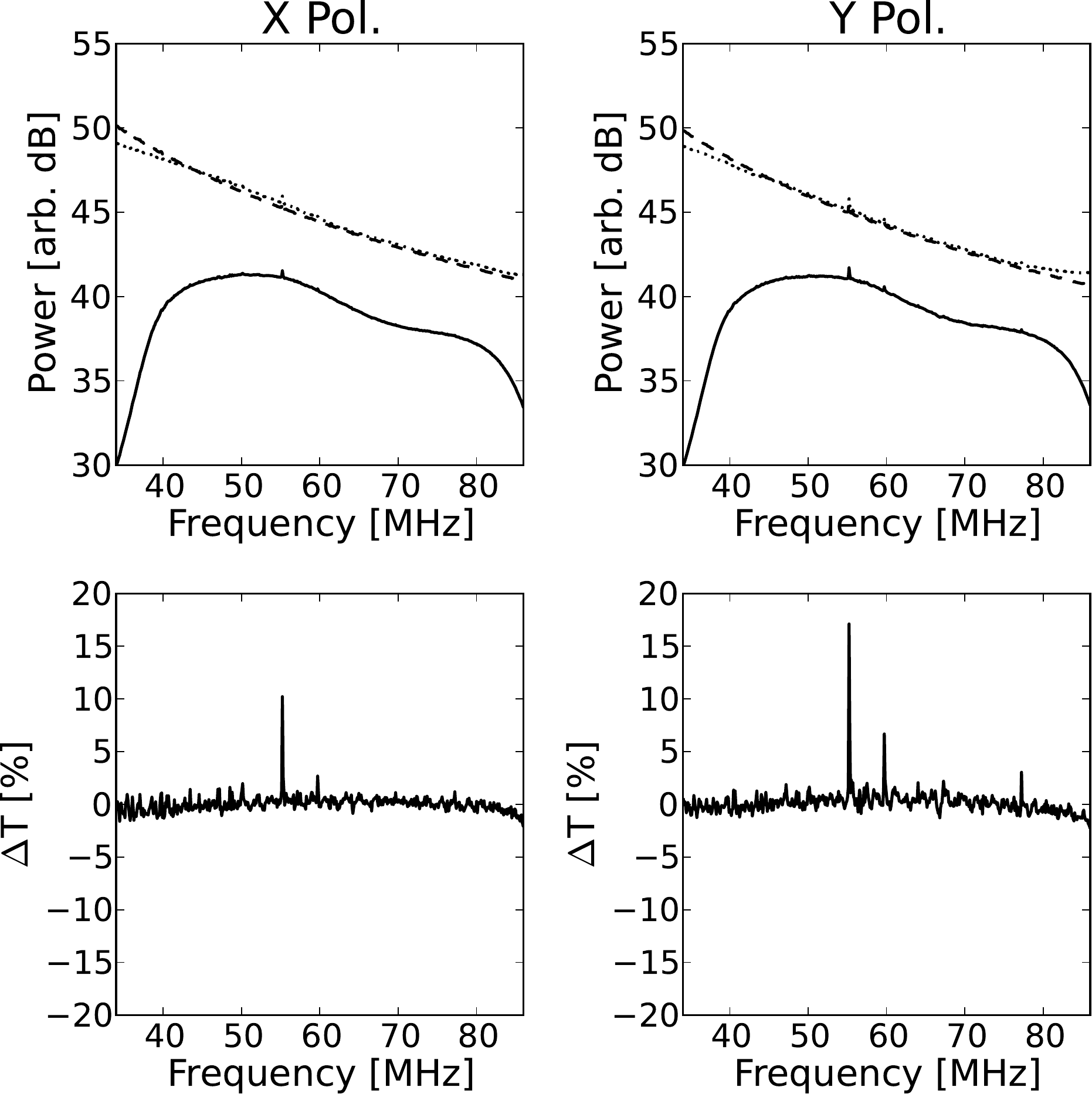}
	\caption{\label{fig:corrections}Corrections to the raw TBW autocorrelations for the antenna impedance mis-match and the ARX filter response.  The top panel series shows the effect of the instrumental corrections (dotted lines) and residuals-based corrections (dashed lines) on the power spectral density as a function of frequency.  The raw spectra (solid lines) are a median composite of eight TBW captures over the local sidereal time range of 12 to 13 hours.  The lower series of plots shows the percentage difference between the observed temperature and a realisation of the Global Sky Model at the same sidereal time that has been convolved with the antenna dipole gain pattern of an isolated antenna.  Overall, the two types of corrections provide a reasonable match to the GSM realisation at the few percent level.}
\end{figure}

After the data were correlated, the visibility amplitudes were corrected for the estimated antenna impedance mis-match and the measured amplitude responses of both the front-end electronics and the analog signal processor\footnote{No additional correction for the receiver noise temperature was applied since the noise temperature is $\sim$250 K and spectrally flat across the frequency range of LWA1 \citep{Henning2010}.}.  The impedance mis-match correction was estimated from the electromagnetic simulations of the dipole antenna design carried out by \citet{AntZ}.  The results of these corrections are shown in Figure \ref{fig:corrections} for the mean X and Y autocorrelation spectra for one of the TBW captures.  The latter two corrections adequately remove the overall bandpass shape and bring the spectra into $\approx\pm$5\% agreement with a model of the sky made at the same sidereal time convolved with the dipole gain pattern.  However, below $\approx$45 MHz and above $\approx$85 MHz the corrected bandpasses begin to show significant deviation, on the order of 10 to 20\%.  This is likely due to limitations of the electromagnetic models of the antenna impedance mis-match or frequency-dependent ground losses.  To correct for this we make the assumption that the average sky signal is spectrally simple and can be well modelled by two terms:  a spectral index and a spectral curvature.  Using this we fit a power law with curvature to the corrected spectra over the frequency range of 45 to 80 MHz and then use this to remove any residual deviations or ripples in the data.  Once this residuals-based correction has been applied, the scatter in the difference between the GSM realisation of the sky and the scaled spectra drops to the few percent level at all frequency bands.  Thus, we estimate that there is a $\sim$5\% uncertainty in the final flux calibration due to uncertainties in the amplitude corrections.  It should also be noted that this correction does not explicitly assume a value for the spectral index or curvature nor does it explicitly tie the data to the GSM or a particular flux density scale.

\subsection{Delay Calibration}
Although the LWA1 archived cable and system delays were applied as part of the correlation process described above, the actual delays for the observations may be slightly different due to variations in the cable lengths as a result of seasonal temperature changes.  Thus, images with a better dynamic range may be possible if a new delay calibration is found using the survey data.  Using TBW captures taken within one hour of either side of the Cygnus A transit, we derived a new delay calibration for each day of data.  These captures were re-phased to Cygnus A and co-added to improve the per-channel signal to noise.  Baselines shorter than 30$\lambda$ at 49 MHz were removed in order to filter out the diffuse emission and the Galactic plane.  The calibration was then carried out using the LSL {\tt lsl.imaging.selfCal} module over 35 to 85 MHz using a model of the sky containing Cassiopeia A, Cygnus A, Hercules A, Sagittarius A, and the Sun.  The delays determined from this step were then used to update the phases of the complex visibilities.  This procedure resulted in a few percent improvement in the dynamic range of the images.  This small improvement in the image fidelity is not surprising given that the archival delays were determined in early 2013 March, less than one month before the first set of TBW captures.  No additional calibration for the ionosphere was applied since the expected position shifts due to refraction at these frequencies are much smaller than the beam size.

\subsection{Imaging and Deconvolution}
\label{sec:img}
\begin{figure}
	\plotone{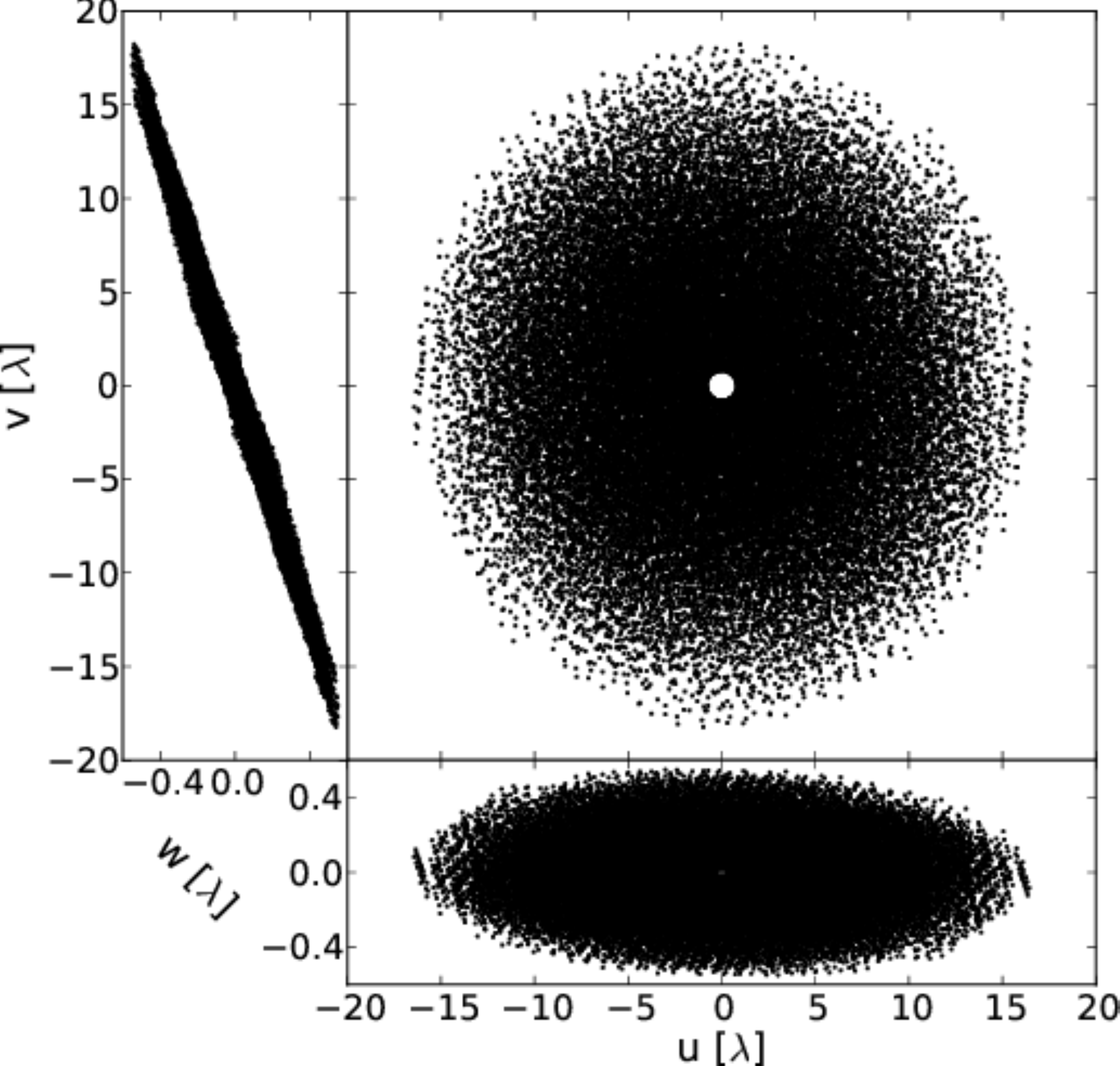}
	\caption{\label{fig:uvCoverage}$(u,v)$ coverage plot for a 50 MHz zenith snapshot at LWA1.  The main panel shows the $(u,v)$ distribution, while the panels flanking the left and bottom sides provide the $(v,w)$ and $(u,w)$ profiles, respectively.  LWA1 was designed to deliver excellent snapshot coverage for all-sky imaging.}
\end{figure}

Imaging and deconvolution were performed using the {\tt wsclean} wide-field imaging software of \citet{wsclean}.  {\tt wsclean} implements a $w$-stacking algorithm with a variety of CLEAN methods and is specifically designed for the case of all-sky imaging.  For all frequencies we generated naturally weighted maps for the XX and YY polarisations separately and deconvolved the maps with multi-scale CLEAN.  Natural weighting was chosen over other weighting schemes since each snapshot is marginally confusion noise dominated and any loss in sensitivity would degrade the image quality.  Each map is 350 pixels square with a central (zenith) pixel size of 20$\arcmin$~in an orthographic sine projection.  The deconvolution parameters, e.g., major loop gain, threshold, and multi-scale bias, were optimized for each frequency individually and were chosen such that the final images were cleaned down to the estimated confusion noise limit.  In all cases the deconvolution halted before the iteration limit was reached.  Table \ref{tab:confusion} gives the frequencies and bandwidths imaged.  In order to regularise the $(u,v)$ coverage and synthesised beam we excluded baselines involving the five outrigger antennas.  In addition, 29 antennas were removed that showed high noise levels that are attributed to noise pickup along the signal paths on the analog signal processor boards.  These two cuts reduced the number of baselines imaged down to 21,115.  Figure \ref{fig:uvCoverage} shows the $(u,v)$ coverage for a snapshot zenith image at 50 MHz.  Even after these cuts the $(u,v)$ plane is well filled down to an inner radius of $\approx$0.8$\lambda$.

\subsection{Flux Calibration}
\label{sec:flx}
In order to create a consistent calibration for the maps that is tied to an existing flux density scale, we have used the following three part strategy.  First, as described in Section \ref{sec:obs} and in Figure \ref{fig:corrections}, we applied broadband corrections to the complex visibilities in order to correct for the antenna impedance mis-match losses and for the bandpasses of the front end electronics and analog receiver chain.  Next, the deconvolved snapshot images were corrected for the dipole gain pattern in order to remove any position dependent response introduced by the antenna.  This correction for the gain pattern was done using an electromagnetic model of the LWA dipole antenna \citep{Dipole1,Dipole2} combined with an empirical elevation-based correction to that model derived from observations of bright pulsars.  For the empirical component we observed seven bright pulsars (PSR B0329+54, B0823+26, B0834+06, B1508+55, B1839+56, B1919+21, and B2217+47) at a variety of elevations ranging from the horizon to upper culmination.  We then estimated the dipole gain by subtracting the off-pulse power from the peak on-pulse power\footnote{Although interstellar scintillation causes the apparent flux density of pulsars to change over time, we do not expect this to impact these observations due to the timescales ($\lesssim$10 hr), bandwidths ($\sim$1 MHz), and observing frequencies (<100 MHz) involved.  Furthermore, the results from the individual pulsars were averaged which down weights the potential impact of scintillation on any one pulsar.}.  From this we found good agreement between the electromagnetic models and the observations above an elevation of 30$^\circ$.  Below this, however, we find that the model underestimates the gain pattern by an amount that increased with decreasing elevation.

These three corrections removed the large-scale frequency variations across the maps.  The next step was to find a scale value that converts the arbitrary units of the map into physically meaningful units.  In order to do this, we compared the observed flux density of Cygnus A with that determined by \citet{Baars}.  Since the $\geq$2$^\circ$ resolution of the imaging does not resolve out the diffuse emission of the Galactic plane around Cygnus A, we used a simple local ``sky subtraction" procedure, analogous to that used for aperture photometry at optical wavelengths, to isolate Cygnus A. For this we selected an annulus around Cygnus A with an inner radius of 1.5 times the beam full width at half maximum (FWHM) and outer radius of 2.5 times the FHWM.  The median of the pixels in the ring was then used to establish the local diffuse background level which was subtracted from a Gaussian fit to the inner region to determine the flux of Cygnus A.  Since Cygnus A is not visible at all sidereal times we calculated on-the-fly flux ratios with Cygnus A for Cassiopeia A, Virgo A, and Taurus A to form a sequence of secondary calibrators.  Table \ref{tab:calibrators} provides flux ratios with Cygnus A for the three secondary calibrators along with the predicted ratios from \citet{Baars}.  Overall the observed ratios are roughly comparable to those of \citet{Baars}, and the differences are not surprising given the difference in resolution between our measurements and those referenced in the Baars scale.  Although the background subtraction of our data is good at removing large scale features, the low resolution of our data means that the subtraction process is less accurate in highly structured regions, particularly for sources that lie near Galactic emission features.  Furthermore, Cassiopeia A displays an unusually high uncertainty in the ratio relative to the other two secondary calibrators.  This variability is likely due to the fact that Cassiopeia A is circumpolar at the latitude of LWA1 and can be observed at lower elevations longer than Virgo A or Taurus A.  Thus, the observed ratio is more sensitive to projection effects and dipole gain pattern errors that are more prevalent at lower elevations.  The final flux scale for a given frequency is determined by using the median flux scale for each source measured in each image.  Combining the uncertainty in the various amplitude corrections applied and the flux density scale factors we estimate an uncertainty in the flux density calibration of 20\%.  This includes uncertainties in both the secondary calibration sources as well as residual errors in the dipole gain pattern.

\begin{table*}
	\centering
	\caption{Secondary Flux Calibrator Ratios to Cygnus A}
	\label{tab:calibrators}
	\begin{tabular}{lcccccc}
		\hline
		Frequency & \multicolumn{2}{c}{Cassiopeia A} & \multicolumn{2}{c}{Virgo A} & \multicolumn{2}{c}{Taurus A} \\
		(MHz) & Observed & Expected$^{a,b}$ & Observed & Expected$^a$ & Observed & Expected$^a$ \\
		\hline
		35 & 1.29 $\pm$ 0.57 & 1.07 & 0.15 $\pm$ 0.05 & 0.20 & 0.11 $\pm$ 0.02 & 0.11 \\
		38 & 1.26 $\pm$ 0.43 & 1.05 & 0.15 $\pm$ 0.04 & 0.19 & 0.10 $\pm$ 0.02 & 0.11 \\
		40 & 1.19 $\pm$ 0.38 & 1.03 & 0.14 $\pm$ 0.03 & 0.19 & 0.10 $\pm$ 0.02 & 0.12 \\
		45 & 1.16 $\pm$ 0.30 & 1.01 & 0.14 $\pm$ 0.03 & 0.18 & 0.10 $\pm$ 0.02 & 0.12 \\
		50 & 1.13 $\pm$ 0.24 & 0.99 & 0.13 $\pm$ 0.02 & 0.18 & 0.10 $\pm$ 0.02 & 0.12 \\
		60 & 1.10 $\pm$ 0.21 & 0.95 & 0.13 $\pm$ 0.02 & 0.17 & 0.10 $\pm$ 0.02 & 0.13 \\
		70 & 1.08 $\pm$ 0.22 & 0.93 & 0.12 $\pm$ 0.03 & 0.16 & 0.10 $\pm$ 0.02 & 0.13 \\
		74 & 1.07 $\pm$ 0.23 & 0.92 & 0.12 $\pm$ 0.03 & 0.16 & 0.10 $\pm$ 0.02 & 0.13 \\
		80 & 1.08 $\pm$ 0.24 & 0.91 & 0.12 $\pm$ 0.02 & 0.15 & 0.10 $\pm$ 0.03 & 0.14 \\
		\hline
		\multicolumn{7}{p{.8\textwidth}}{$^a$  Calculated from \citet{Baars}.}\\
		\multicolumn{7}{p{.8\textwidth}}{$^b$  Assumes a secular decrease of 0.77\% per year \citep{CasA}.} \\
	\end{tabular}
\end{table*}

\subsection{Conversion to Temperature and Missing Spacings Correction}
Next the snapshot images were converted from intensity to temperature using the two dimensional beam area.  The motivation for this conversion is two-fold.  First, the primary motivation of this survey is to provide a more complete picture of foregrounds in the low frequency sky for 21-cm cosmology experiments.  Since these experiments deal with the temperature of the spin-flip transition, converting the sky maps to temperature facilitates a more direct comparison for these experiments.  Second, as will be explained below, the conversion to temperature also makes the correction for the missing spacings in the snapshots more straightforward.

After the snapshots had been converted to temperature, the correction for the missing spacings was determined.  Although the minimum LWA1 antenna spacing of five meters provides a relatively small inner hole in the $(u,v)$ plane (0.6$\lambda$ at 35 MHz and 1.2$\lambda$ at 74 MHz) there is a considerable amount of Galactic emission on the largest spatial scales missing from each snapshot image.  This emission is particularly important for 21-cm cosmology applications that seek to make a total power measurement of the hydrogen spin-flip transition.  To correct for the missing emission in the interferometric maps we use data from the total power system that is part of the LEDA-64 New Mexico deployment \citep[][hereafter LEDA]{FL} along with forward modelling of LWA1 to determine which spatial scales were missing from each snapshot image.  Briefly, LEDA uses the five LWA1 outriggers to make a total power measurement of the sky through a dedicated recording system.  These five outriggers are equipped with special front end electronics that include a three-state temperature calibration system that can be used to provide the absolute sky-averaged temperature to better than 10 K, or less than 1\%, over 40 to 80 MHz.  Using data taken from a 24-hour LEDA run from 2014 December 6 we corrected for the missing emission as follows.  First, we used the calibrated LEDA total power data to determine a global scale factor to convert the LWA1 mean auto-correlation to temperature.  Since the frequency range of interest for this survey extends beyond the frequency coverage of LEDA, we used the 72 to 76 MHz region where the LEDA impedance mis-match losses are better understood.  Using this reduced spectral coverage is acceptable since we have already removed the relative spectral response of the instrument from the auto-correlations following the procedure outlined in Section \ref{sec:flx}.  With the auto-correlations calibrated we estimated the total missing emission in each linear polarisation by subtracting the sky-averaged temperature in each snapshot from the value expected from the auto-correlations.  

In order to estimate which spatial scales are missing in each snapshot we have used forward modelling of the array to examine the flux recovery at a variety of spatial scales.  For this we used a sky model along with the dipole gain pattern to simulate visibility data for each local sidereal time.  The resulting visibilities can then be imaged and deconvolved to determine which spatial scales are missing from the simulated images by comparing them with the input model.  Since we are seeking a self-consistent survey of the low frequency sky we have derived our input sky model directly from our data.  To help compensate for the missing spacings we have used multi-frequency synthesis \citep{mfs1,mfs2} in {\tt wsclean} to jointly image our data between 28 and 85 MHz.  By imaging data across this frequency range we sample the $(u,v)$ plane from $\lesssim$0.5 $\lambda$ up to $\approx$30 $\lambda$.  Thus, the model includes both detailed spatial structure information from the highest frequencies, while the lowest frequencies are least affected by missing spacings.  In the absence of mutual coupling corrupting the measurements on the shortest baselines, we assume that this approach allows us to recover the total emission from the sky.  Electromagnetic simulations of the full LWA1 array by \citet{mutual} suggest that mutual coupling, while significant, does not adversely impact the performance of the array.  However, we do note that any limitations in this assumption should manifest on large spatial scales ($\gtrsim$100$^\circ$).  Furthermore, the input sky model for our forward modelling resulting from this process shows large-scale structure that is qualitatively similar to what is seen in the GSM.  After projecting this input sky model onto the observed hemisphere for each snapshot and simulating the visibilities, the data were then imaged and calibrated using the methods described in Sections \ref{sec:img} and \ref{sec:flx}, respectively.  The calibrated simulation snapshots were then compared with the input model to create a collection of missing spacing template images for each snapshot, frequency, and polarisation combination\footnote{Due to modelling artefacts around bright sources such as Cygnus A, these regions were replaced with a tilted plane fit to an annulus around each source.}.  These templates were then used to iteratively add large-scale emission to each snapshot until the total power was in agreement with the calibration auto-correlation data.  It should also be noted that although we make an assumption about the nature of the large-scale Galactic emission, this does not necessarily translate to a pre-defined structure for the large scale emission.  Rather, by applying the correction on a per-image and per-polarisation basis we are able to reconstruct some of the large-scale two dimensional structure as the sky rotates through the field of view.

\subsection{Mosaicking}
\label{sec:mos}
Once each image was deconvolved and calibrated, the resulting XX and YY images were upsampled using bilinear interpolation and then combined to Stokes I.  The Stokes I snapshots were then re-projected onto a HEALPix RING grid \citep{HEALpix} with 256 sides, or an approximate pixel size of one-quarter of a degree.  This resolution is well matched to the $\approx$0.33 degree pixel resolution of the images.  At this stage three image-based cuts were also applied.  The first cut masked the sky below $\approx$16$^\circ$ elevation to remove sources of radio frequency interference (RFI) located along the horizon, while the second removed a region three FWHM in diameter, 14$^\circ$ at 35 MHz and 6$^\circ$ at 80 MHz, around the location of the Sun.  The region around the Sun is filled in the final mosaicked maps by the two TBW runs taken roughly a month apart.  Finally, each image was visually inspected to remove any residual RFI or images that showed deconvolution artefacts.  After flagging, the images were averaged together using a robust (outlier-resistant) method to generate the sky maps.

\section{Results}
\label{sec:res}
\begin{figure*}
	\epsscale{2.2}
	\plotone{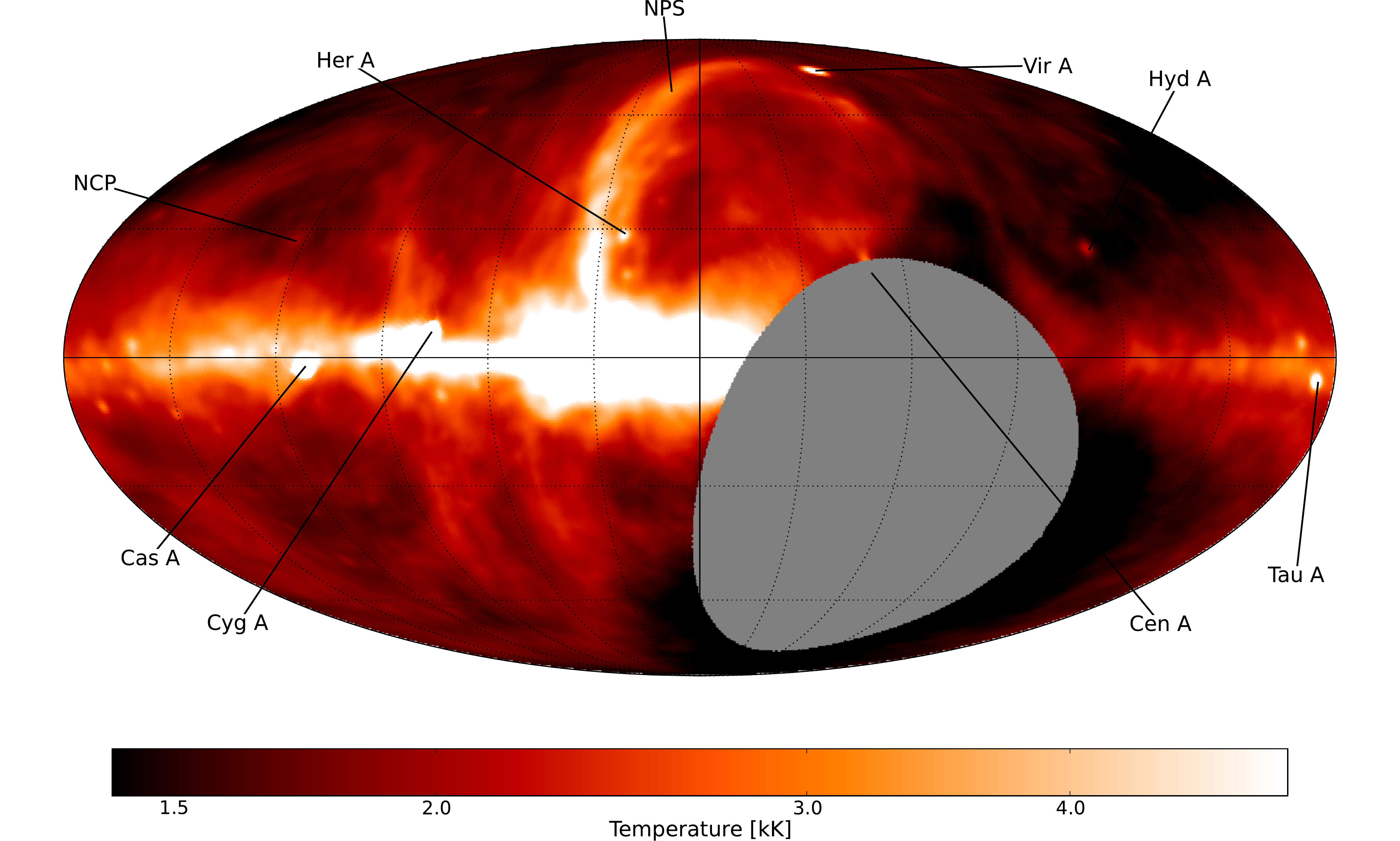}
	\caption{\label{fig:mapLarge}74 MHz map imaged over $\approx$1 MHz of bandwidth.  The map is displayed with a Mollweide projection in Galactic coordinates with latitude and longitude marked every 30$^\circ$.  The colour scaling is logarithmic and the beam size at this frequency is 2.2$^\circ$.  Prominent sources, such as Cygnus A and Cassiopeia A are labeled, along with the north celestial pole (NCP) and the north polar spur (NPS).  The systematically lower temperatures near the south Galactic pole are likely a result of limitations in the dipole response model and the corrections for the missing spacings.}
\end{figure*}

\begin{figure*}
	\epsscale{1.6}
	\plotone{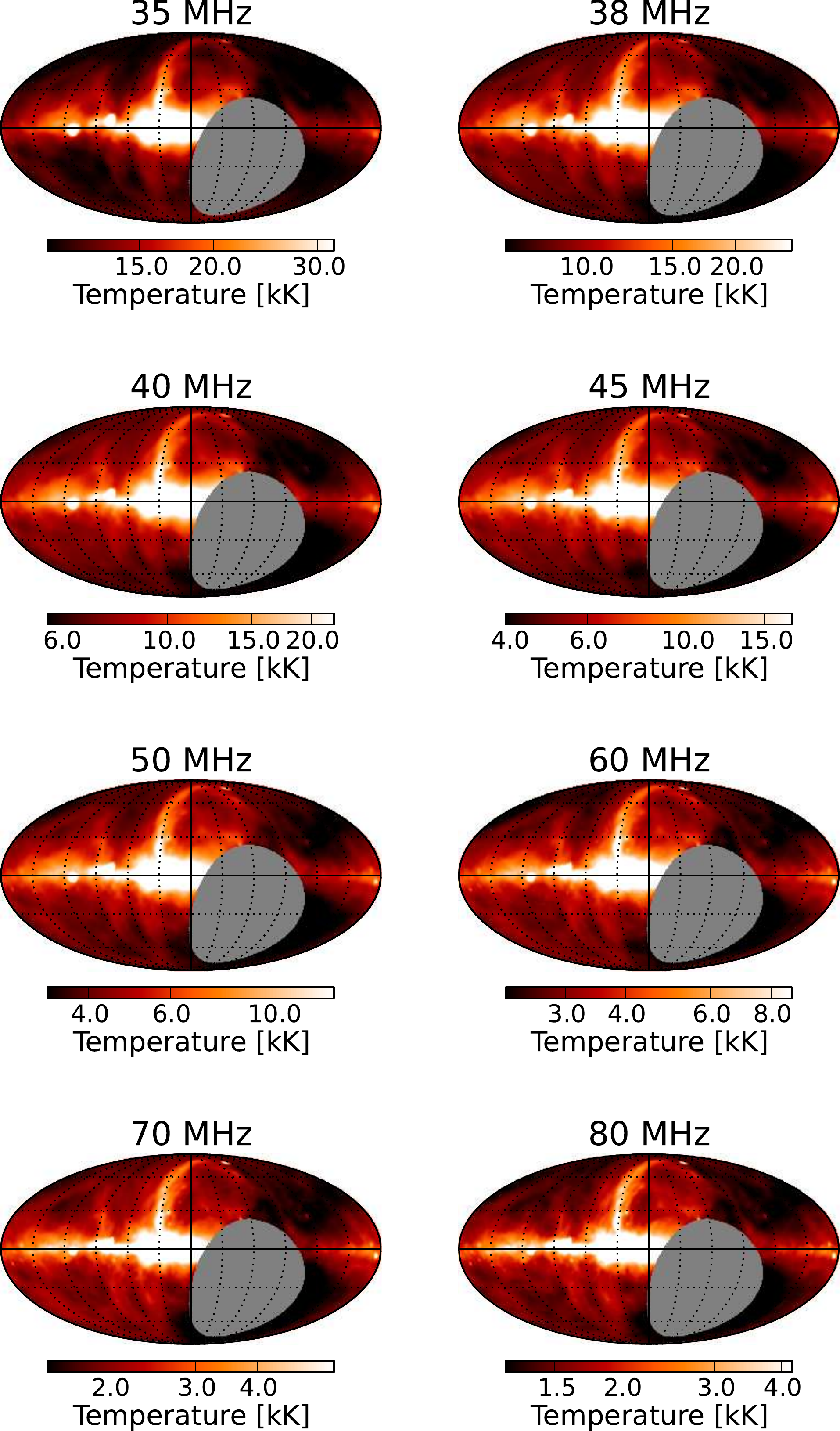}
	\caption{\label{fig:mapSet}Maps at 35, 38, 40, 45, 50, 60, 70, and 80 MHz imaged over $\approx$1 MHz of bandwidth.   The layout is the same as in Figure \ref{fig:mapLarge} except that the source labels have been omitted.  The beam size varies from 4.7$^\circ$ at 35 MHz to 2.0$^\circ$ at 80 MHz.}
\end{figure*}

Figures \ref{fig:mapLarge} and \ref{fig:mapSet} show the deconvolved and calibrated maps in Galactic coordinates for all nine frequencies:  35, 38, 40, 45, 50, 60, 70, 74, and 80 MHz.  These frequencies were chosen to span the frequency range accessible with the LWA1 in the split bandwidth mode with a regular 10 MHz frequency spacing.  Additionally, we have also included maps at 38 and 74 MHz that correspond to the radio astronomy frequency allocations, and maps at 35 and 45 MHz for comparison with existing data.  The full width at half maximum beam size in the images ranges from about 4.7$^\circ$ at 35 MHz to 2.0$^\circ$ at 80 MHz.  At the lower frequencies the images are dominated by the diffuse emission of the Milky Way, with many of the fainter sources masked by this emission and the relatively large beam size.  As the frequency increases the diffuse emission of the Galaxy begins to weaken and fainter sources, such as Virgo A and Taurus A, become more prominent.  The maps also show a detection of the northern lobe of Centaurus A across all bands.  At the upper end of the band the falling sensitivity of the instrument begins to manifest itself as increased noise.  We also note that the area around the south Galactic pole in all nine maps shows a systematic bias toward lower temperatures which is likely a result of limitations in the model for the dipole response at low elevations coupled with the method used to correct for the missing spacings.  The HEALPix maps in RING format and equatorial coordinates, as well as standard FITS images in a Mollweide projection, are publicly available for download from the LWA Data archive at \url{http://lda10g.alliance.unm.edu/LWA1LowFrequencySkySurvey/}.

As discussed in Section \ref{sec:obs} the individual snapshots, and hence the final maps, should be confusion limited.  To test this we looked at the dynamic range in our final maps.  We measured the ratio between the peak intensity at Cygnus A and a root-mean-square values of 12 blank, ``confusion dominated" areas of the sky distributed in a ring 10$^\circ$ from 3C295.  Although this type of comparison is limited because it cannot disentangle the confusion from the system noise it does provide a straight forward method of evaluating the quality of an individual map.  At 74 MHz we find a ratio of 630$\pm$150, close to the expected value for a confusion-limited image of 790.  At 38 MHz the ratio is 300$\pm$150 which exceeds the expected value of 190 by roughly a factor of two.  The high uncertainties in these values are a reflection of variations in the Galactic foreground emission in this area.

For a more complete look at whether or not the images were confusion limited, we used a modified version of the method proposed by \citet{GSM35}.  For this we first estimated the observed total noise, which is a combination of confusion and thermal noise, in a map by differencing the value at each location with the median value in an annulus one FWHM away and then took the standard deviation of this difference over 20$^\circ$ areas of the sky.  We then estimated the contribution of the thermal noise by differencing maps at two closely spaced frequencies and performing the same analysis as for the total noise.  Since this method relies on having two nearby frequencies we have limited our analysis to the 38/40 MHz map pair and the 70/74 MHz map pair.  In addition, we have only examined regions off the Galactic plane since the confusion noise is likely to be higher in and around the plane due to the structure of the Galaxy.  Table \ref{tab:confusionLimit} shows the results of this analysis over two regions of the sky: one from an RA of 12$^h$30$^m$ to 13$^h$30$^m$ and a declination of $-$30$^\circ$ to $+$70$^\circ$ and the other from an RA of 1$^h$30$^m$ to 2$^h$30$^m$ over the same declination range.  The values for both regions are consistent with the theoretical estimates from Table \ref{tab:confusion} given the difficulty of estimating both quantities from low resolution data and the presence of the north polar spur in the region centred on 13$^h$.  We do note, however, that the thermal noise is roughly a factor of two higher than the theoretical estimates near the top of the LWA1 observing band which may indicate that there are deconvolution artefacts and/or low-level RFI at the $\approx$5 Jy beam$^{-1}$ ($\approx$15 K) level.  Indeed, during the period when the data were taken there was broadband interference from micro-arching on nearby power lines at the LWA1 site.  We also note that this difference is significantly less ($\lesssim$2\%) than the expected Galactic foregrounds at these frequencies.

\begin{table*}
	\centering
	\caption{Confusion and Thermal Noise Estimates}
	\label{tab:confusionLimit}
	\begin{tabular}{lcccccc}
		\hline
		Frequency & \multicolumn{3}{c}{1$^h$30$^m$ to 2$^h$30$^m$} & \multicolumn{3}{c}{12$^h$30$^m$ to 13$^h$30$^m$} \\
		~ & Total Noise & Thermal Noise & Calculated Confusion & Total Noise & Thermal Noise & Calculated Confusion \\
		(MHz) & (Jy beam$^{-1}$) & (Jy beam$^{-1}$) & Noise$^a$ (Jy beam$^{-1}$) & (Jy beam$^{-1}$) & (Jy beam$^{-1}$) & Noise$^a$ (Jy beam$^{-1}$)\\
		\hline
		38 & 42 & 28 & 32 & 105 & 27 & 101 \\
		40 & 39 & 28 & 27 &   90 & 27 & 86 \\
                 70 & 12 & 11 &   5 &  18 & 11 & 14 \\
                 74 & 13 & 11 &   7 &  17 & 11 & 13 \\
      		\hline
		\multicolumn{7}{p{.8\textwidth}}{$^a$  Calculated assuming that the thermal and confusion noise are independent and add in quadrature.}\\
		\end{tabular}
\end{table*}

In addition to the HEALPix maps presented here we have also created two collections of visualisations suitable for the general public.  The first is a Google Maps-style interface which consists of tiles rendered in a Mercator projection.  This interface includes additional frequencies from the Google Sky\footnote{\url{http://www.google.com/sky/}} and can be accessed at \url{http://fornax.phys.unm.edu/}  We also have assembled a set of spherical visualisations based on visualisations of the Planck data at \url{http://thecmb.org} created by \cite{thecmb.org}.  These maps are interactively rendered in a web browser using WebGL and provide a projection of the sky onto a three dimensional sphere.

\begin{figure}
	\epsscale{1.0}
	\plotone{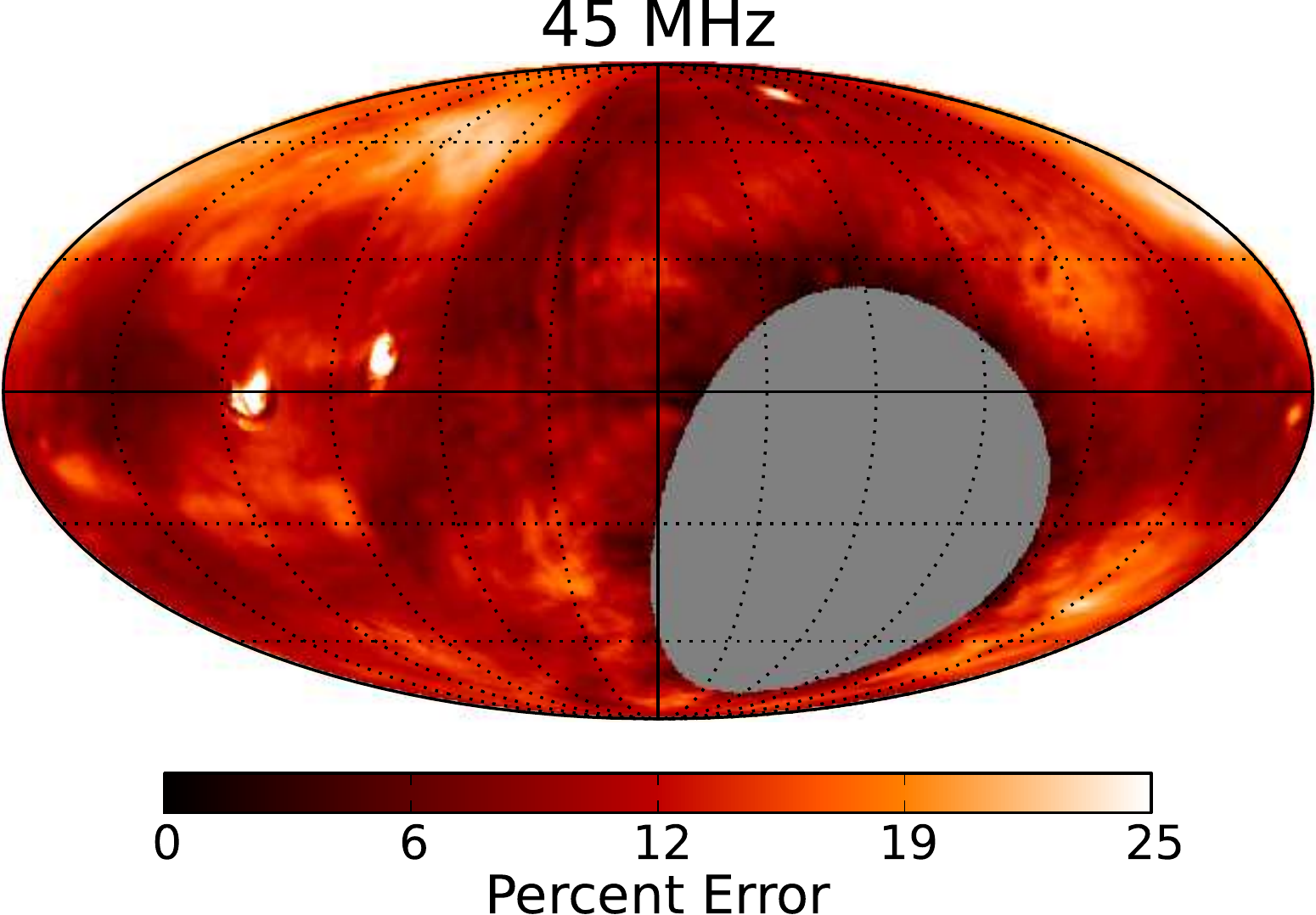}
	\caption{\label{fig:mapError}Uncertainty map for 45 MHz presented as a percentage of the map temperature.  The uncertainty map is in Galactic coordinates and the upper end of the colour scale saturates at 25\%.  The uncertainty is fairly uniform except for higher uncertainties around residual RFI.  Bright sources also show higher error which is likely due to scintillation in the snapshots.}
\end{figure}

The mosaicking method used here also makes it possible to generate per-pixel uncertainty maps as part of the mosaicking process.  For this we examined the per-pixel standard deviation of all snapshots that contribute to that particular pixel.  Thus, the uncertainty map provides an estimate of how well the snapshots agree and captures the effects of uncertainties in the various calibrations and corrections applied to the data.  Figure \ref{fig:mapError} shows a representative uncertainty map at 45 MHz.  Although the overall uncertainty at 45 MHz is on the order of 10\%,  there are several large areas in the map where the uncertainty approaches or exceeds 20\%.  These areas are located between Virgo A and Taurus A, between the north celestial pole and the north polar spur, and near the south Galactic pole.  There are also smaller areas, both scattered around the map and centred on bright sources, which have higher uncertainties.  The larger areas are probably the result of errors in correcting for the missing spacings, while the smaller areas are from RFI that was not masked or ionospheric scintillation.

\section{Analysis}
\label{sec:an}
As indicated in the introduction there are a variety of uses for these maps.  In this section we explore a few of the possible science cases, including how the maps relate to existing sky surveys and models, and what the background spectral index is across the sky at our frequencies.

\subsection{Comparison with Other Maps and Models}
\label{sec:reldiff}
\begin{figure}
	\plotone{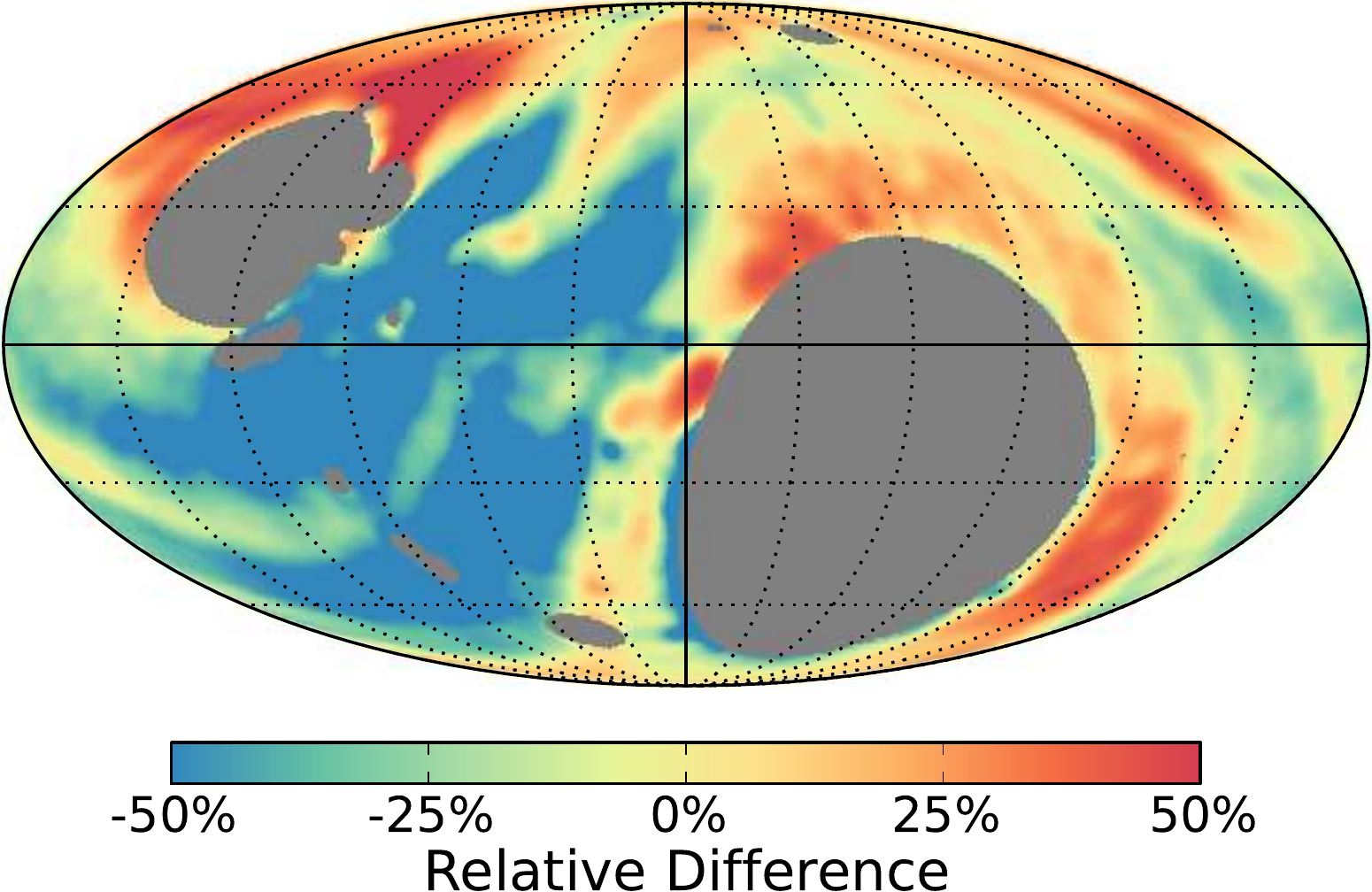}
	\caption{\label{fig:comp35}Comparison between our 35 MHz map and the 34.5 MHz map of \citet{GSM35} for areas where the maps overlap.  The comparison is done by subtracting the comparison map from our sky map and then dividing by our sky map.  The colour scale is saturated at $\pm$50\% to provide a overall sense of the differences and $+$50\% (red) corresponds to more emission in our map.  In general the 35 MHz map presented here shows less emission overall by $\approx$20\%.}
\end{figure}

Within the compendium of maps compiled by \citet{GSM} for building the GSM, only two within the frequency range considered here provide sufficient areal coverage and resolution for comparison:  the 34.5 MHz map of \citet{GSM35} and the 45 MHz map of \citet{GSM45A} and \citet{GSM45M}.  We compare our map at 35 MHz with the resolution-matched 34.5 MHz map by taking the difference between the two maps and dividing that difference by our map.  The resulting ratio map shown in Figure \ref{fig:comp35} is dominated by two features.  The first and most significant difference between the maps is that our map shows $\approx$20\% less emission, with most of the discrepancy being in the area south of the Galactic plane.  The second major feature is a collection of three red bands which correspond to lines of constant declination.  These lines are at roughly --20$^\circ$, 20$^\circ$, and 60$^\circ$ that correspond to likely artefacts noted by \citet{GSM35} in their map.  The source of the difference in the area south of the plane is unclear but we do note that our map is consistent with our 45 and 74 MHz maps at the few percent level.

\begin{figure}
	\plotone{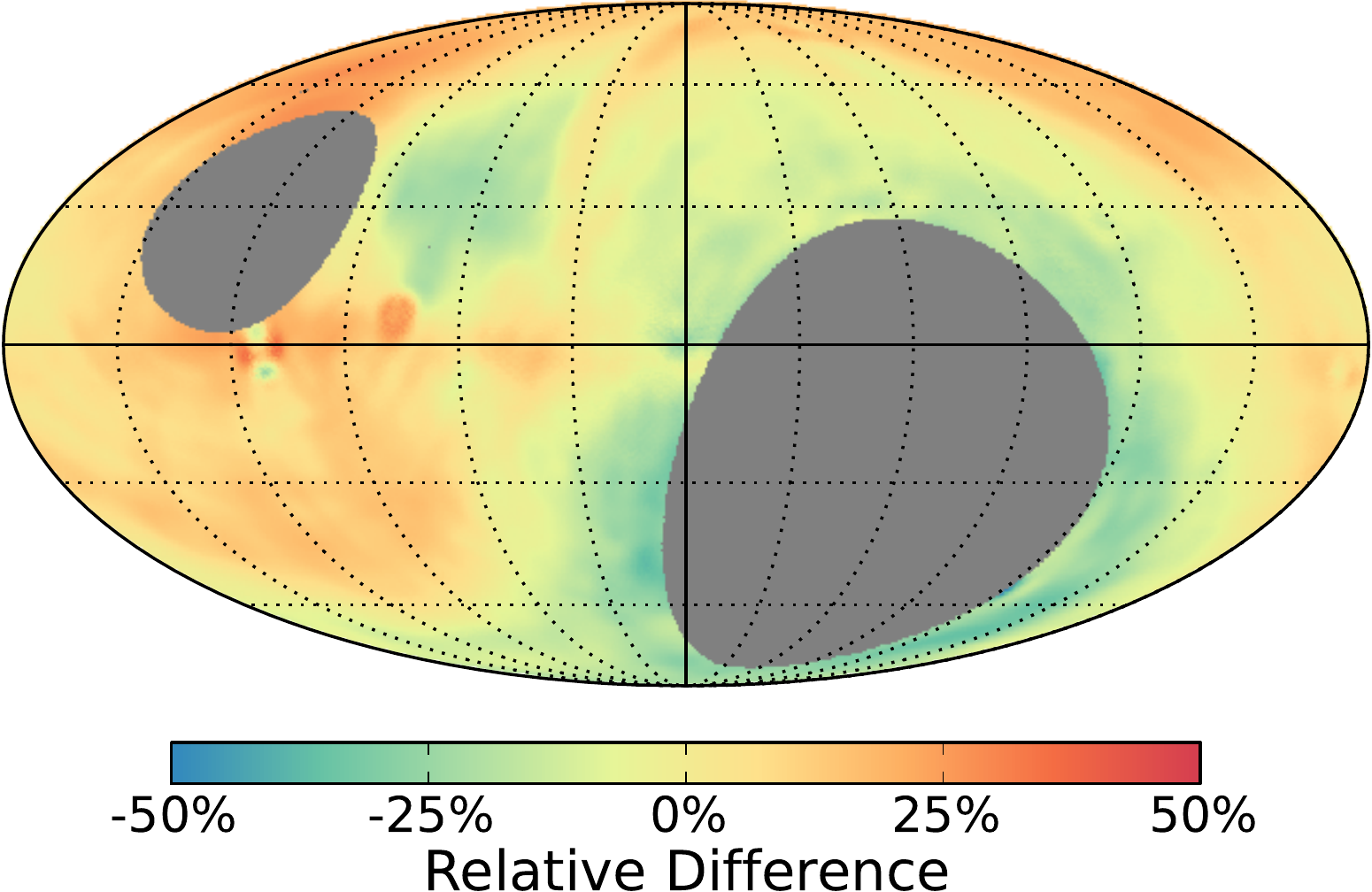}
	\caption{\label{fig:comp45}Comparison between our map and the map of \citet{GSM45A} and \citet{GSM45M} at 45 MHz for regions where the maps overlap.  The colour scale is the same as in Figure \ref{fig:comp35}.  The 45 MHz map of this work shows more emission in the Galactic plane than the map of \citet{GSM45A} and \citet{GSM45M}.}
\end{figure}

\begin{figure}
	\plotone{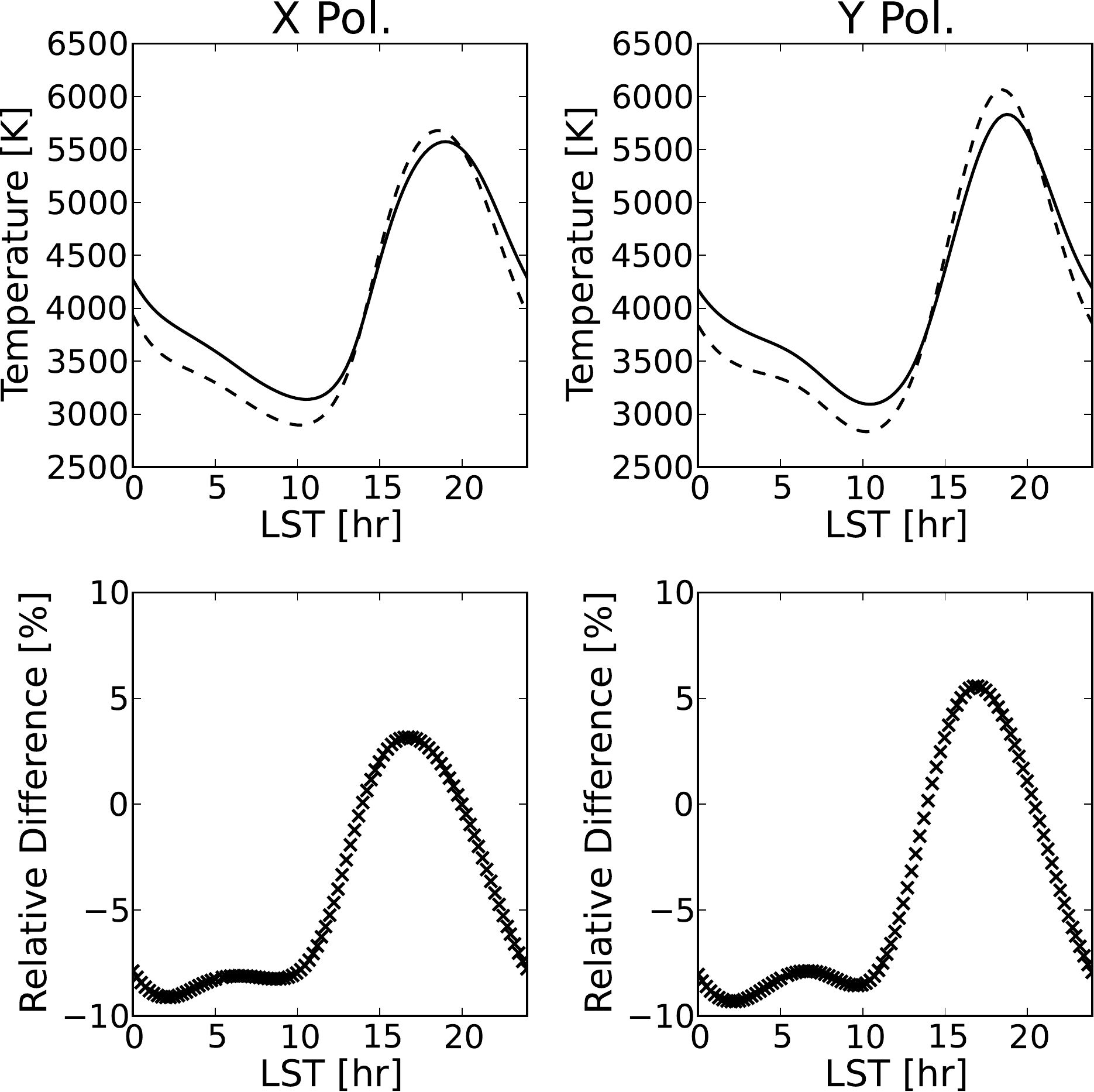}
	\caption{\label{fig:comp45DC}Comparison of simulated drifts at 45 MHz of our maps (solid line) and the 45 MHz map of \citet{GSM45A} and \citet{GSM45M} (dashed line).  The lower panels show the difference between the drift curves as a fraction of our temperature where negative values indicate that the LWA1 drift curve has a higher temperature.  Overall there is agreement between the maps at the 10\% level.}
\end{figure}

Figure \ref{fig:comp45} shows the ratio between our 45 MHz map and that of \citet{GSM45A} and \citet{GSM45M}.  The comparison here shows better overall agreement than the 35 MHz comparison, with the sky-averaged temperatures being within 2\% of each other.  The two largest areas of difference are the region between Virgo A and Taurus A, and near the south Galactic pole.  These two regions appear as orange and blue in the figure, respectively.  To examine how these differences influence the total power for experiments such as LEDA, we have simulated drift curves for the X and Y polarisations.  The drift curves are created by projecting the maps onto the sky at various sidereal times and convolving the result with a model of the dipole gain pattern.   Figure \ref{fig:comp45DC} shows the results of this comparison, and we find that both maps are consistent at the $\approx$10\% level, with our map generally showing a higher temperature.

\begin{figure}
	\plotone{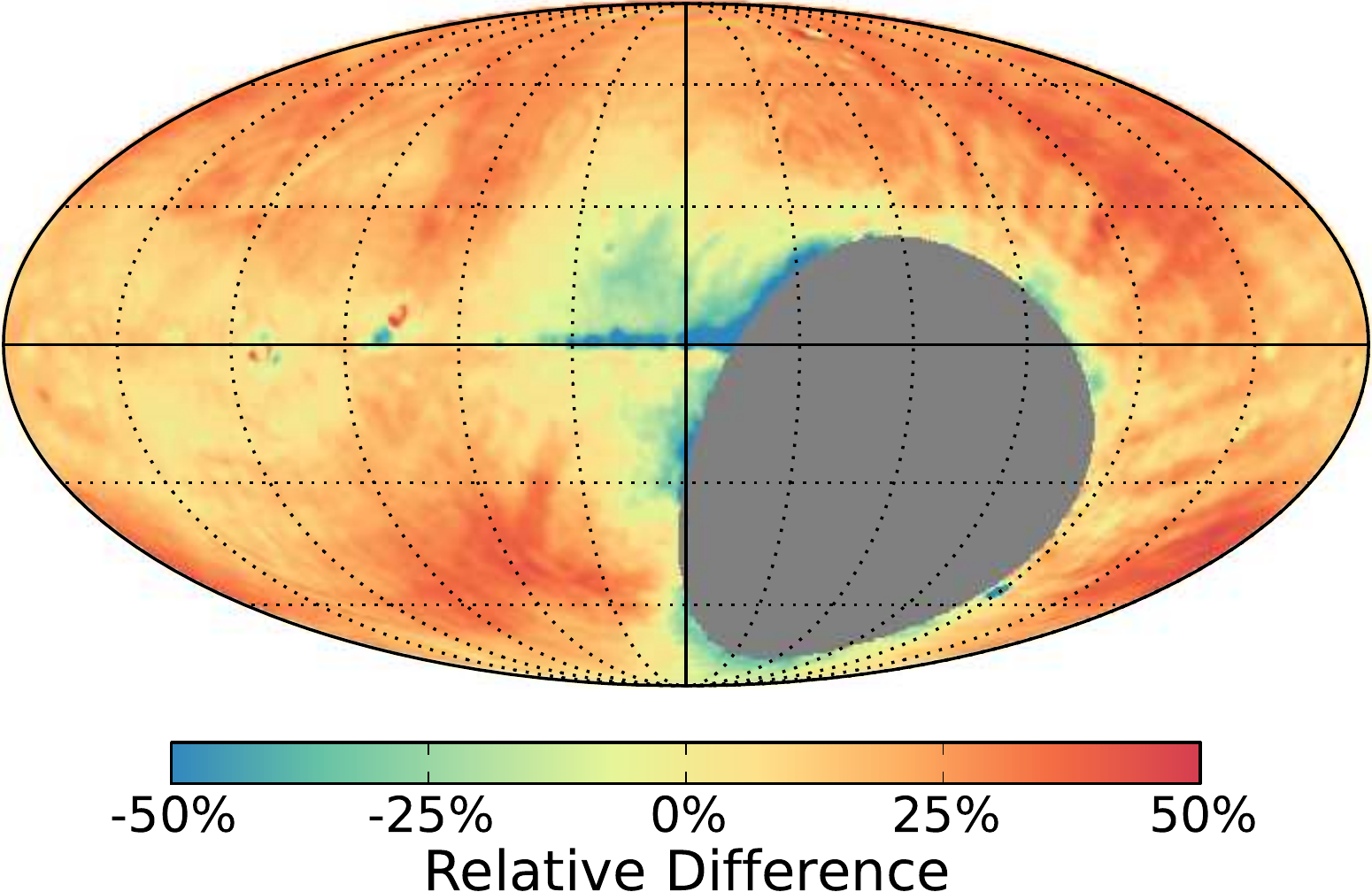}
	\caption{\label{fig:comp74}Comparison between our 74 MHz map and a 74 MHz realisation of the GSM.  The colour scale is the same as in Figure \ref{fig:comp35}.  Overall there is agreement between the two maps at the 20\% level.}
\end{figure}

\begin{figure}
	\plotone{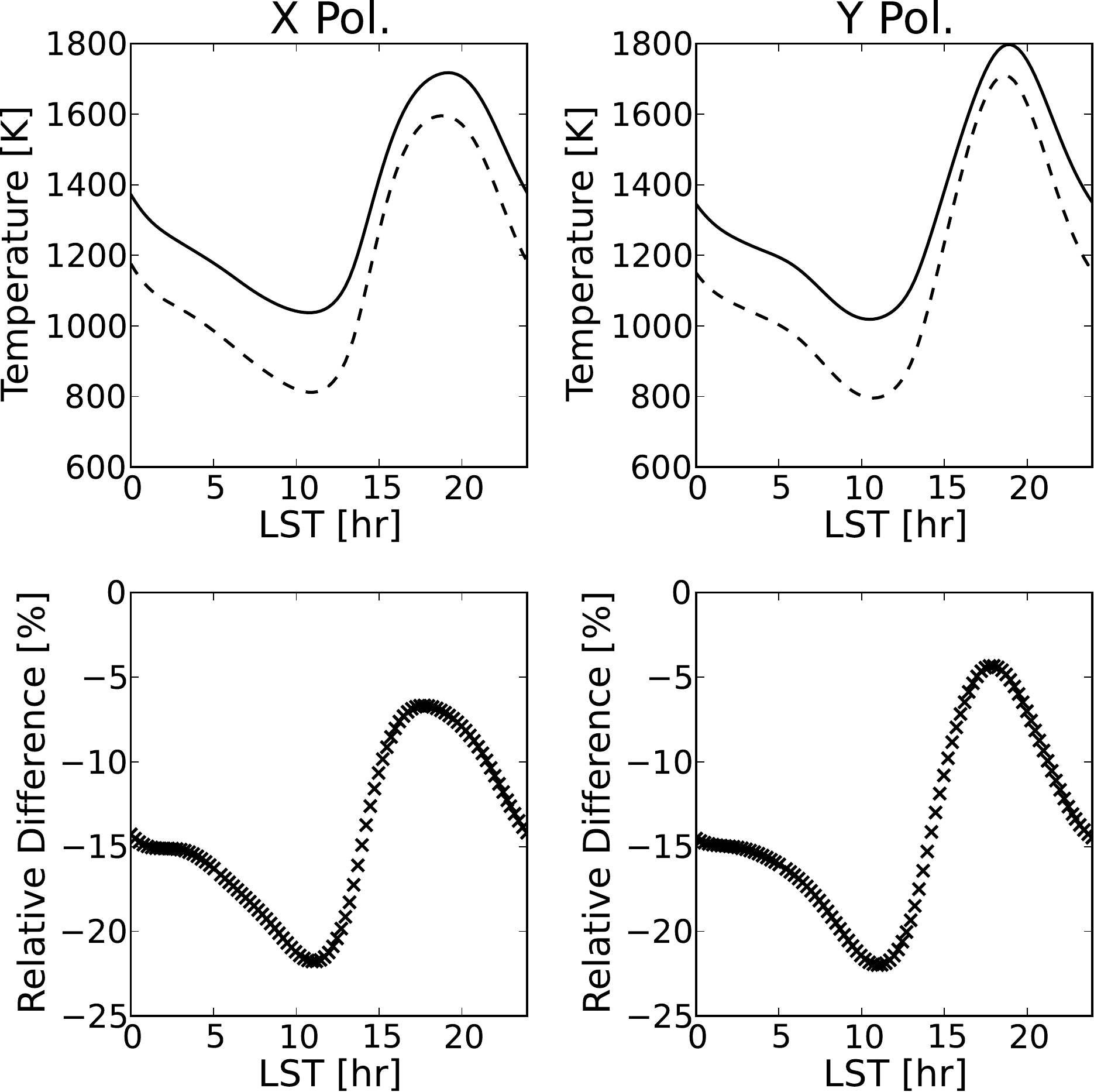}
	\caption{\label{fig:comp74DC}Comparison of simulated drifts at 74 MHz of our maps (solid line) and a realisation of the GSM at 74 MHz (dashed line).  The layout is the same as in Figure \ref{fig:comp45DC}.  There is good agreement in the structure of the drifts although we find a systematic offset between our map and the GSM with our map having temperatures $\approx$16\% higher.}
\end{figure}

In addition to comparisons with individual maps it is also possible to compare against the GSM.  This is of particular interest given that the majority of the frequencies that go into the GSM analysis are above 1 GHz.  Below this only four maps are used:  10, 22, 45, and 408 MHz \citep{GSM}.  Thus, the GSM in the LWA1 frequency range relies heavily on the higher frequency maps.  In Figure \ref{fig:comp74} we compare our 74 MHz map with a resolution-matched GSM realisation at the same frequency.  Overall there is good agreement both in the spatial structure and the flux scale between the two maps, with our map having about a 15\% higher temperature than the GSM realisation.  This offset is more clearly shown in the total power drift curve simulation presented in Figure \ref{fig:comp74DC}.  Similar to 45 MHz, the areas with a systematically lower temperature are present although the region near the south Galactic pole is smaller.  There also appears to be residual stripping between Virgo A and Taurus A from combining the individual snapshots.  Figure \ref{fig:comp74} also shows about 30\% less emission along the Galactic plane, particularly near the Galactic centre, relative to the GSM.  This is likely due to the effects of free-free absorption around \ion{H}{2} regions in the plane that are poorly constrained by the higher frequencies that dominate the GSM.

\subsection{Spectral Index Maps}
\label{sec:spindex}
\begin{figure}
	\plotone{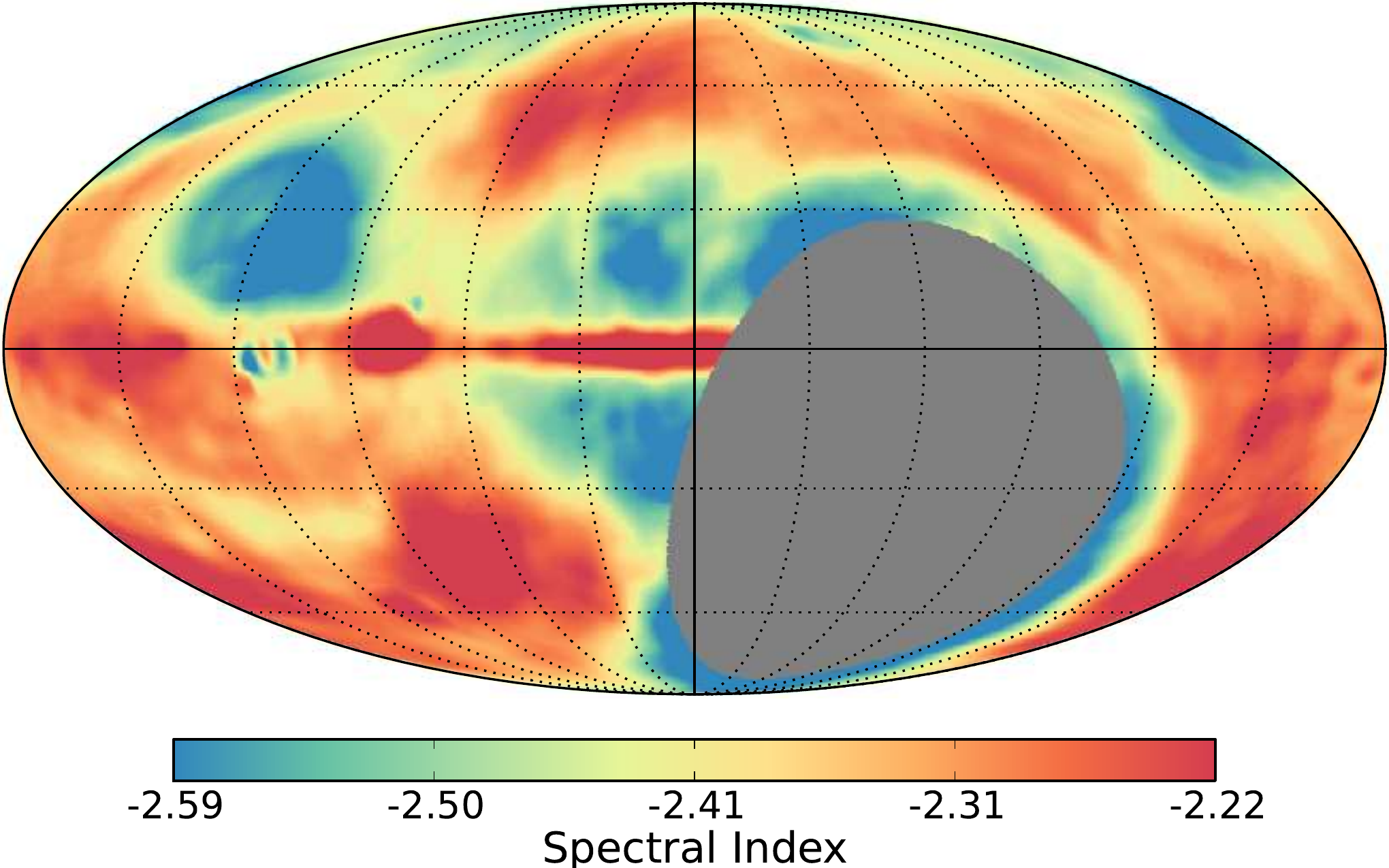}
	\plotone{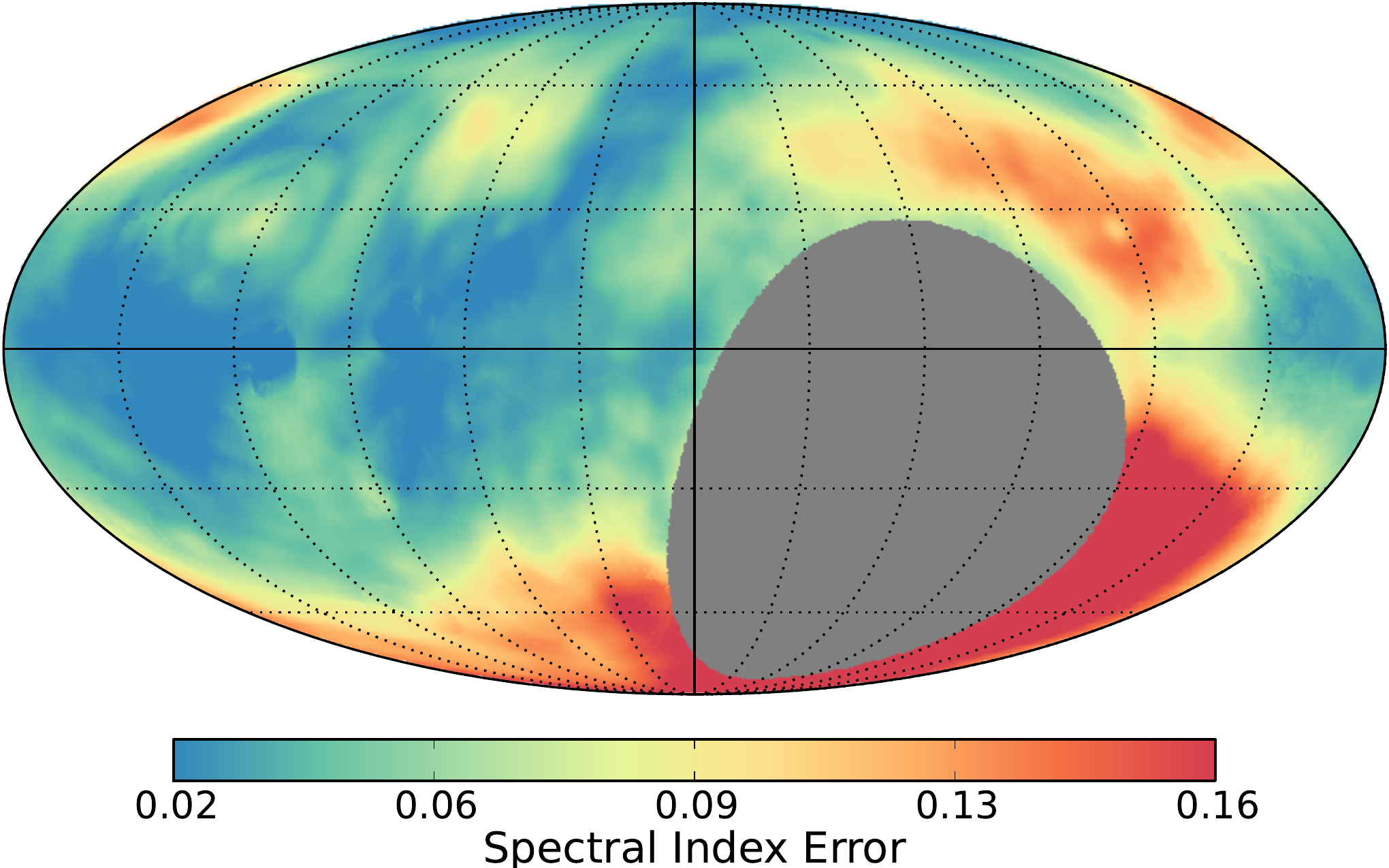}
	\caption{\label{fig:spindex}Spectral index map (top) and the estimated uncertainty (bottom) computed by combining all nine maps.  The colour scale is displayed such that steeper features are blue and shallower features are red.  The sky is relatively smooth spectrally with a flattening in the Galactic plane which is consistent with the spectral index map of \citet{Guzman11}.  There is also an area near the north celestial pole that shows an unusually steep spectrum.  In addition, the area near the south Galactic pole has an unusually shallow spectrum.  Both of these are likely spurious  artefacts arising from deconvolution artefacts or problems correcting for the missing spacings.}
\end{figure}

Figure \ref{fig:spindex} shows a spectral index map of the sky and the associated uncertainty computed exclusively using our nine maps.  The map is dominated by the Galactic plane.  There are two regions that show an unusually shallow spectral index of $\sim$--2 outside of the plane:  one between Virgo A and Hydra A that was noted in the comparison at 74 MHz and another near a Galactic latitude of --60$^\circ$.  These are likely artefacts due to the shallowness of the spectral index and the lack of strong Galactic structure in those areas.  Since the estimated uncertainty in this region is less than 0.1 this feature appears to be a systematic artefact across many frequencies.  There is also an apparent artefact near the north celestial pole where the map shows a steeper spectral index relative to the immediate surroundings.

\begin{figure}
	\plotone{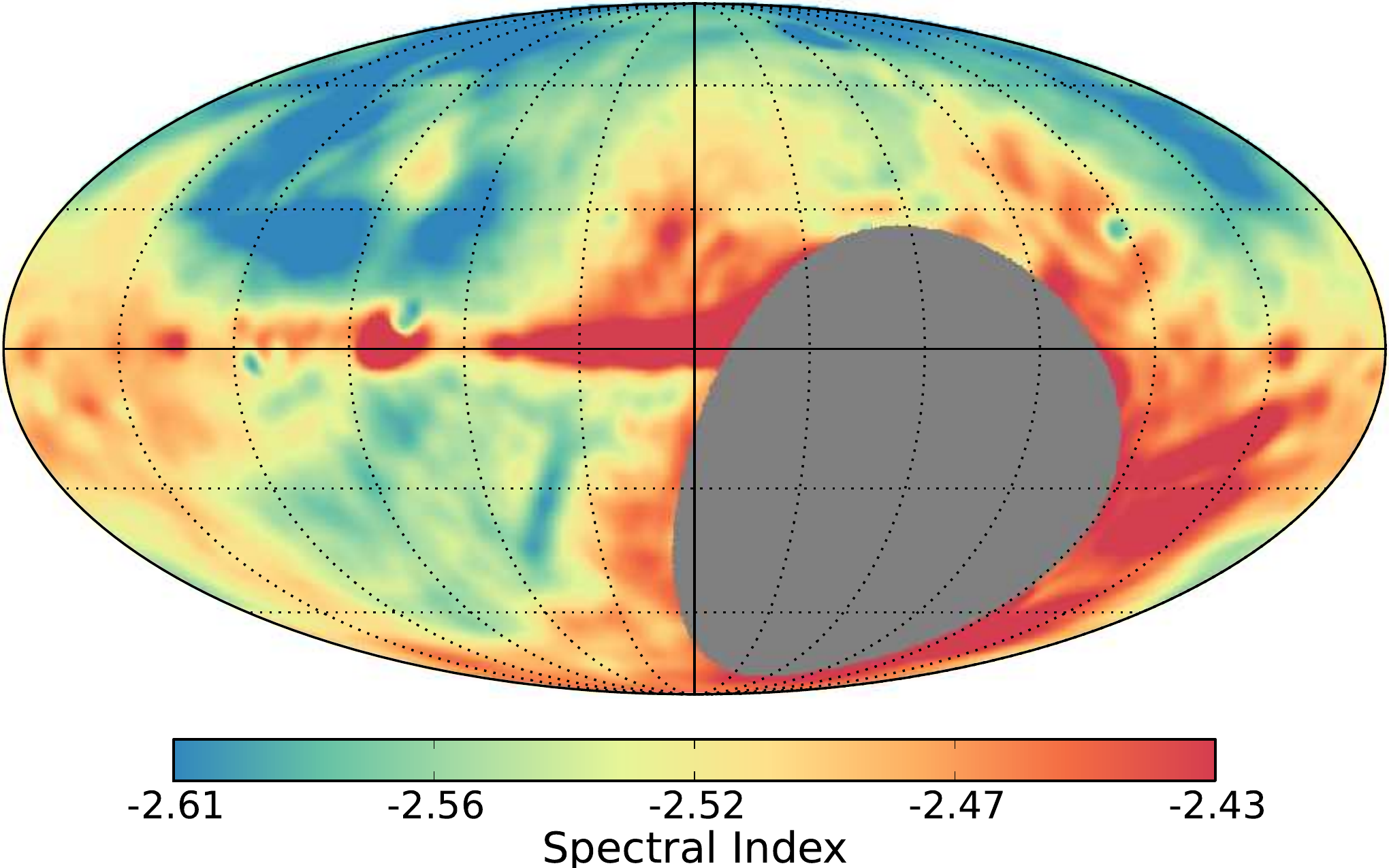}
	\plotone{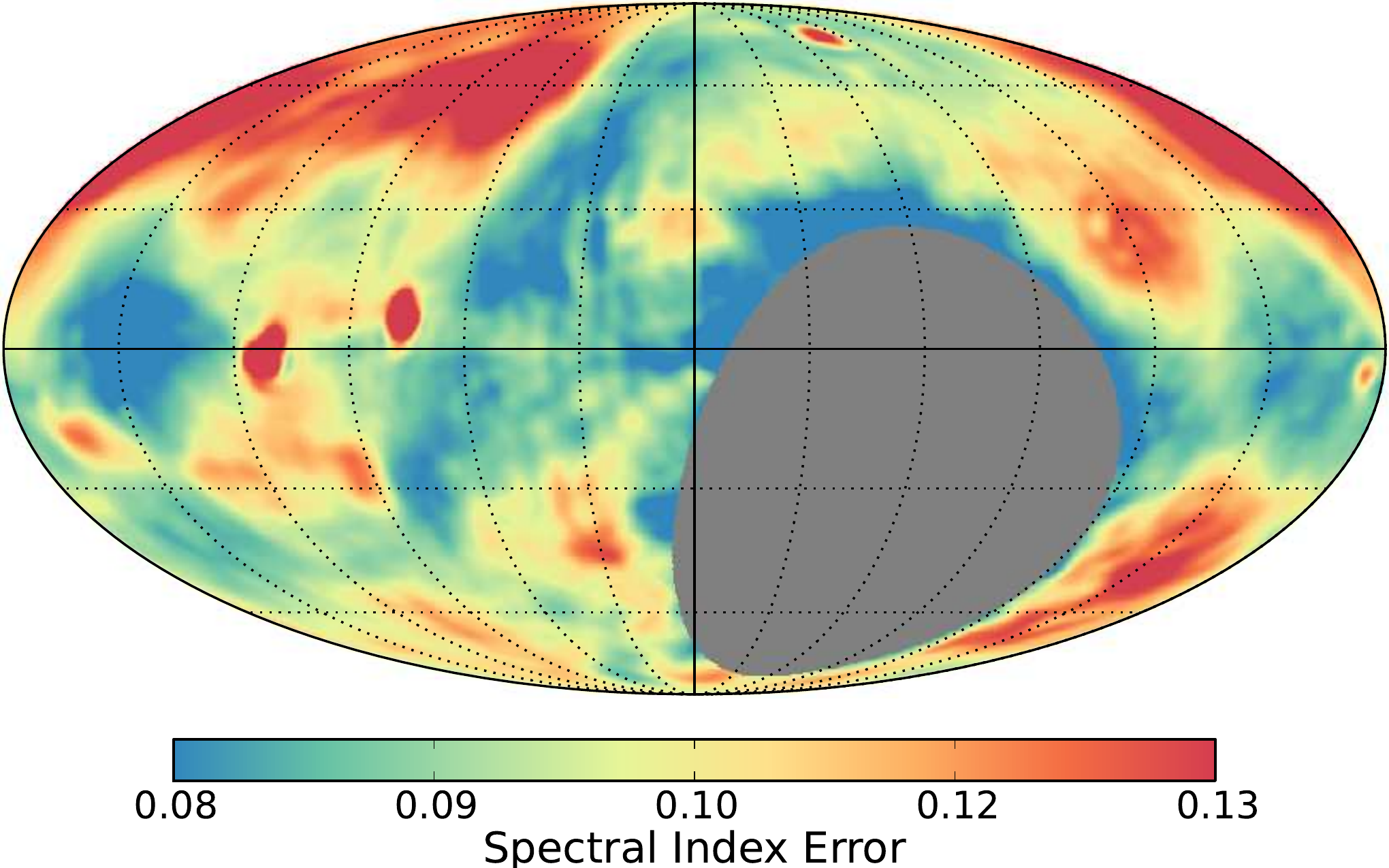}
	\caption{\label{fig:spindexWide}Spectral index map (top) and the estimated uncertainty (bottom) computed between our 45 MHz map and the reprocessed 408 MHz map of \citet{eHaslam}.  The spatial resolution and colour mapping are the same as in Figure \ref{fig:spindex}.  The uncertainties were estimated using the LWA1 error map at 45 MHz and assuming a 10\% uncertainty in the 408 MHz map.}
\end{figure}

Barring the feature near the south Galactic pole, the spectral index map shown in Figure \ref{fig:spindexWide} is similar to what is shown in \citet{Guzman11} and other work at low frequencies.  However, it is important to note that most spectral index maps are computed between a single low frequency and the reprocessed version 408 MHz map of \citet{GSM408} done by \citet{eHaslam}.   Following this approach, Figure \ref{fig:spindexWide} shows the spectral index of the sky calculated between our 45 MHz map and 408 MHz smoothed to the same resolution of 4.7$^\circ$.  The largest difference between this spectral index map and our LWA-only spectral index is in the regions above and below the plane that are more shallow in the LWA-only map.  We also see artefacts with a shallower index near the southern declination limit which could be related to limitations of the dipole gain pattern response at the lowest elevations coupled with uncertainties in the corrections for the missing spacings applied to the data.

Outside of these anomalous regions, many of the differences between the spectral index maps in Figures \ref{fig:spindex} and \ref{fig:spindexWide} could be related to the differences in the assumed spectral structure of the sky at these frequencies.  When using two widely spaced frequencies, like \citet{Guzman11} and others, a power law is the simplest, natural functional form to fit.  However, this approach averages over several underlying emission and absorption mechanisms that cannot be fit by a single power law. In addition to the cosmic microwave background and extragalactic background as discussed in \citet{Guzman11}, there is both non-thermal and thermal Galactic emission at 408 MHz, and non-thermal emission and thermal absorption at the LWA frequencies. The latter is due to \ion{H}{2} regions and their extended envelopes \citep{Kassim89}.  This must result in the deviation from a single power law particularly in regions within the Galactic plane.  Modelling the composite emission spectrum by combining our maps with existing radio surveys at other frequencies should be possible, however, this analysis is beyond the scope of this paper and will be addressed in a future work.

\section{An Updated Model of the Low Frequency Sky}
\label{sec:lfsm}
Since the maps presented here cover $\approx$80\% of the visible sky from 35 to 80 MHz it is now possible to revisit a model for the low frequency sky.  We have constructed an updated model, the Low Frequency Sky Model (LFSM), for use below 400 MHz that follows the modelling approach used for the Global Sky Model of \citet{GSM}.  Briefly, this approach uses a two part principal component analysis of the constituent maps to create a model of the sky that can be interpolated to an arbitrary frequency within the model's range.  The first part of the analysis examines the sky averaged signal over the area covered by all of the input maps in order to capture the overall spectral structure of the sky.  With this sky-averaged signal described, the second part of the analysis is used to explain how the two dimensional structure of the sky changes with frequency.  For further details on the mathematics behind this method see the discussion of \citet{GSM}.

\subsection{Input Survey Maps}
\begin{figure}
	\plotone{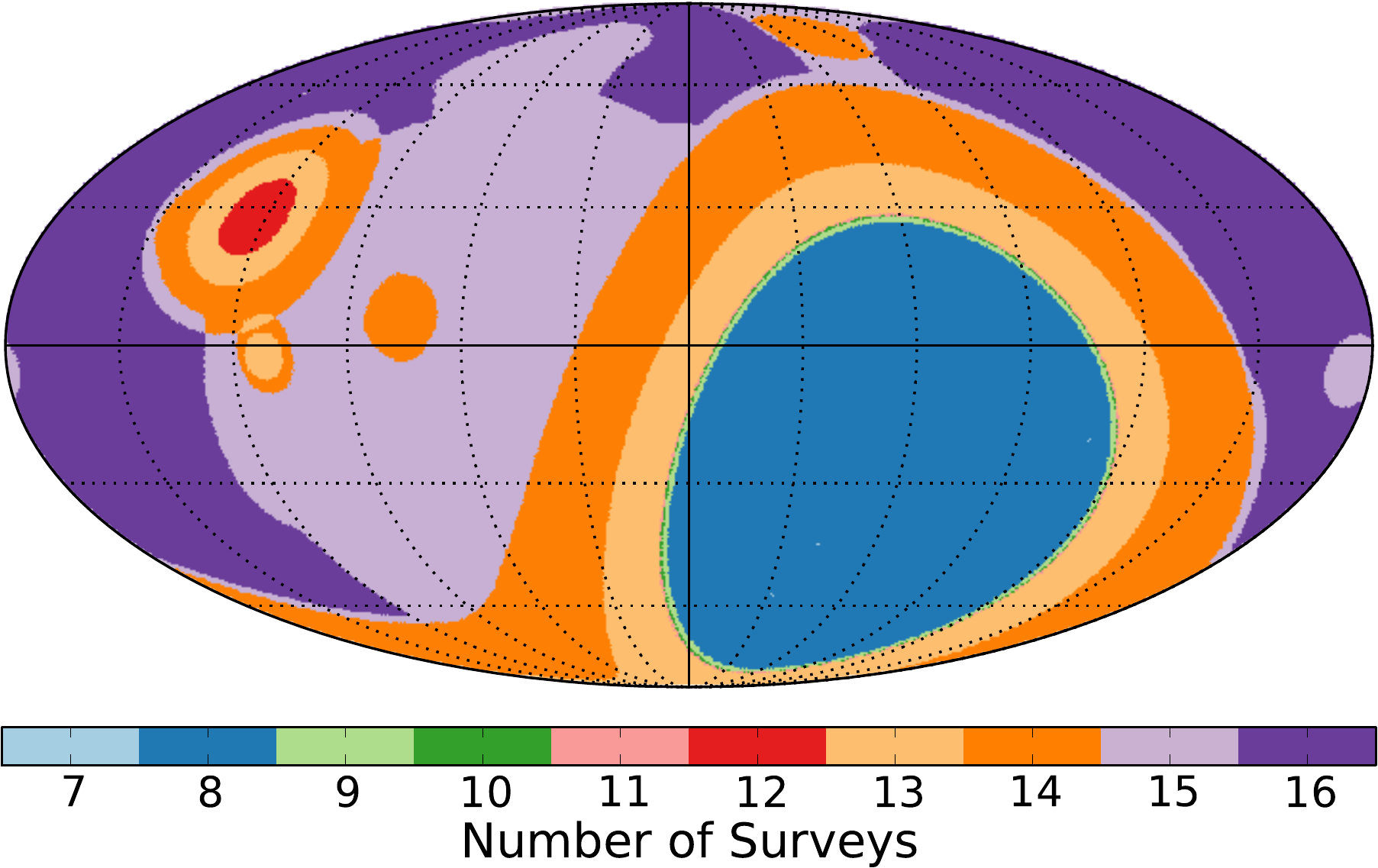}
	\caption{\label{fig:coverage}Sky coverage maps in Galactic coordinates for our updated low frequency sky model.  The colours represent the number of surveys that contribute data at each pixel.}
\end{figure}

The input survey maps to the LFSM are given in Table \ref{tab:inputs}.  From our work we used the regularly spaced maps at 40, 50, 60, 70, and 80 MHz.  These fives maps are combined with 11 maps from the literature to provide full spectral information from 10 MHz to 94 GHz for approximately 24\% of the sky (Figure \ref{fig:coverage}).  From the literature we have also chosen to include the 45 MHz map of \citet{GSM45A} and \citet{GSM45M} rather than our own map at the same frequency because it provides coverage of the region around the south celestial pole.

\begin{table*}
	\centering
	\caption{Input Survey Maps to the Low Frequency Sky Model}
	\label{tab:inputs}
	\begin{tabular}{lcccl}
		\hline
		Frequency & \multicolumn{2}{c}{Coverage} & FWHM &Reference(s) \\
		MHz & RA & Dec & ($^\circ$) &  ~ \\
		\hline
		10       & 0$^h$ < $\alpha$ < 16$^h$ &  --6$^\circ$ < $\delta$ < +74$^\circ$ & 2.6 $\times$ 1.9 & \citet{GSM10} \\
		22       & 0$^h$ < $\alpha$ < 24$^h$ & --28$^\circ$ < $\delta$ < +80$^\circ$ & 1.7 $\times$ 1.1 & \citet{GSM22} \\ 
		40       & 0$^h$ < $\alpha$ < 24$^h$ & --40$^\circ$ < $\delta$ < +90$^\circ$ & 4.3 $\times$ 3.9 & This work \\
		45       & 0$^h$ < $\alpha$ < 24$^h$ & --90$^\circ$ < $\delta$ < +65$^\circ$ & 5.0                      & \citet{GSM45A,GSM45M} \\
		50       & 0$^h$ < $\alpha$ < 24$^h$ & --40$^\circ$ < $\delta$ < +90$^\circ$ & 3.8 $\times$ 3.5 & This work \\ 
		60       & 0$^h$ < $\alpha$ < 24$^h$ & --40$^\circ$ < $\delta$ < +90$^\circ$ & 2.8 $\times$ 2.6 & This work \\
		70       & 0$^h$ < $\alpha$ < 24$^h$ & --40$^\circ$ < $\delta$ < +90$^\circ$ & 2.4 $\times$ 2.2 & This work \\
		80       & 0$^h$ < $\alpha$ < 24$^h$ & --40$^\circ$ < $\delta$ < +90$^\circ$ & 2.1 $\times$ 2.0 & This work \\
		408     & \multicolumn{2}{c}{all sky}                                                                 & 0.8                      & \citet{GSM408,eHaslam} \\
		819     & 0$^h$ < $\alpha$ < 24$^h$ &  --7$^\circ$ < $\delta$ < +85$^\circ$ & 1.2                      & \citet{GSM820} \\
		1419   & \multicolumn{2}{c}{all sky}                                                                 & 0.6                      & \citet{GSM14201,GSM14202,GSM14203} \\
		23000 & \multicolumn{2}{c}{all sky}                                                                 & 0.9                      & \citet{GSM} \\
		33000 & \multicolumn{2}{c}{all sky}                                                                 & 0.7                      & \citet{GSM} \\
		41000 & \multicolumn{2}{c}{all sky}                                                                 & 0.5                      & \citet{GSM} \\
		61000 & \multicolumn{2}{c}{all sky}                                                                 & 0.4                      & \citet{GSM} \\
		94000 & \multicolumn{2}{c}{all sky}                                                                 & 0.2                      & \citet{GSM} \\
		\hline
	\end{tabular}
\end{table*}

\subsection{Component Fitting}
\begin{figure}
	\plotone{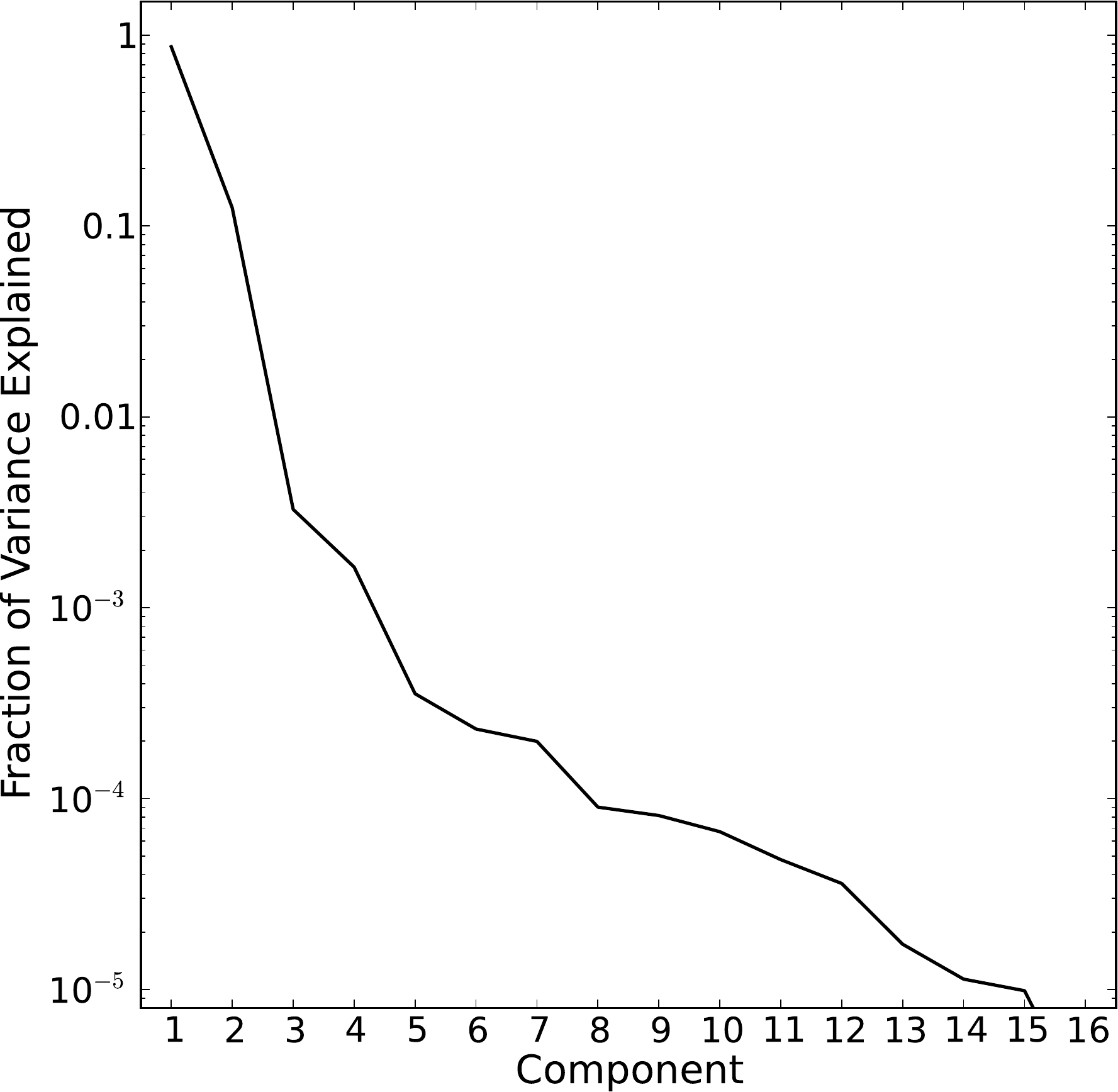}
	\caption{\label{fig:eigen}Normalised eigenvalues from the principal component analysis of the region of the sky that is observed by all ten maps.  Most ($\approx$99.7\%) of the spectral variation of the sky can be described with only three components.}
\end{figure}

The principal component analysis was performed using an eigenvector decomposition of the correlation matrix computed from the region of the sky with full coverage from all 16 maps.  The normalised eigenvalues associated with each of the eigenvectors are shown in Figure \ref{fig:eigen}.  We find that most of the variation with frequency between the maps, $\approx$99.7\%, can be described using only three components.  Although more components can be included to increase the accuracy of the decomposition, we limit the analysis to three components to avoid overfitting.  Once the basis vectors were determined we found three component maps that, when projected onto the eigenvectors, captured the two dimensional structure of the sky.  Since the resolutions of the input maps vary between 0.2$^\circ$ at 94 GHz and 5.0$^\circ$ at 45 MHz we have smoothed the maps to the lower resolution of the 45 MHz map.  The component maps were found by performing a least squares fit to the input survey data on a pixel-by-pixel basis.  Unlike \citet{GSM} we found that the quality of the resulting component maps was strongly dependent on the choice of the noise covariance matrix used for the fitting.  This problem was most noticeable in areas with incomplete spectral coverage and likely resulted from the limited amount of frequency information in those regions.  To overcome this we employed an iterative scheme to estimate the diagonal terms of the noise covariance matrix.  We performed the component map fitting using a subset of eight maps and used the resulting fit to examine the root-mean-square (RMS) difference between the excluded map and the prediction.  Once the RMS difference had been determined for all nine survey maps, we updated the covariance matrix and recomputed the fits.  This continued until the RMS change for all maps dropped below a threshold of 1\% between iterations.  Once the optimal covariance matrix had been determined, the sky was fit at each pixel in the model using all surveys that contributed data to that pixel.  To avoid potential problems caused by any residual uncertainty in the LWA1 dipole gain pattern we apodise the weight of the LWA1 data for declinations less than --10$^\circ$ with a simple quadratic function.  This function smoothly lowers the relative weight of the LWA1 data to zero at the southern declination limit of --40$^\circ$.  After fitting we further smoothed the component maps to a resolution of 5.1$^\circ$ to reduce edge effects at the survey boundaries.

The LFSM is available on the LWA Data Archive at the same location as the LWA1 Low Frequency Sky Survey data.  The model realisation is implemented as a Python script that performs a cubic interpolation between the principal components and then sums over the components.  It should be noted that the realisation software has been implemented so that it also works with the components and component maps of \citet{GSM} via a compatibility mode.

\subsection{Accuracy}
\begin{figure}
	\plotone{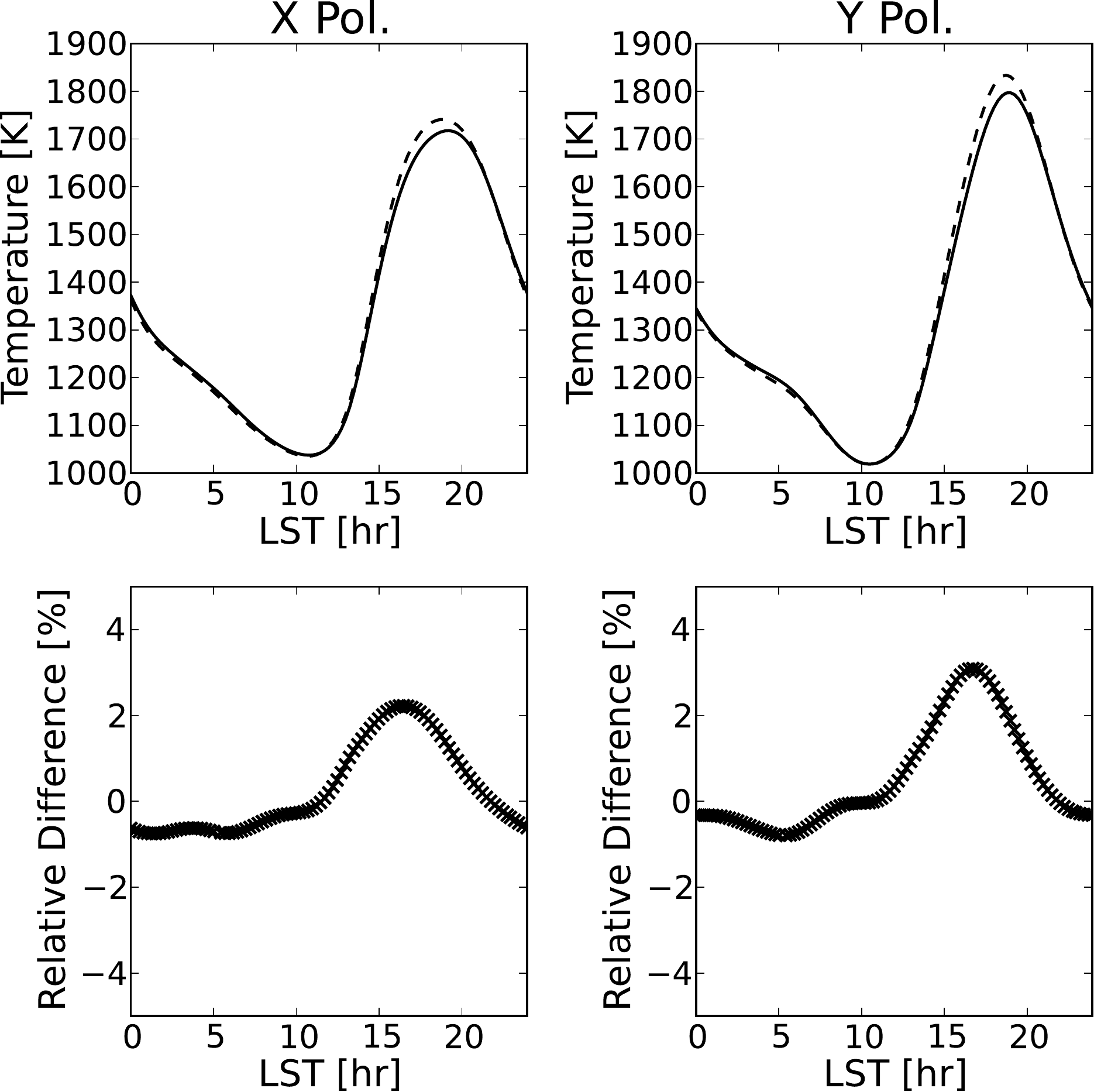}
	\caption{\label{fig:lfsmDrifts}Comparison of simulated drift curves at 74 MHz between our map (solid lines) and a realisation of the LFSM (dashed lines).  The layout is the same as in Figure \ref{fig:comp45DC}.  The low frequency sky model shows agreement with our data at this frequency at the few percent level.}
\end{figure}

To test the accuracy of the model for total power purposes we have created simulated drift curves to compare the model realisations to our map at 74 MHz.  This frequency was chosen because it was not used in generating the model.  The total power comparison is shown in Figure \ref{fig:lfsmDrifts}.  There is agreement between the model realisation at this frequency and our survey at the few percent level.  The largest difference between the realisation of the LFSM and our map occurs around a local sidereal time of about 18 hours, which corresponds to when the Galactic centre is overhead.  Here we find that the realisation over predicts the average temperature by a few percent and this is likely a result of the inclusion of the 408 MHz to 94 GHz data.  Alternatively, it may be the result of having used only three components within the model.  Since the components are determined from large areas of the sky, they capture the overall change of the sky with frequency.  However, this does not necessarily mean that un-modelled components are uniformly distributed over this area and they may be concentrated in a particular area such as the Galactic plane.

\begin{figure}
	\plotone{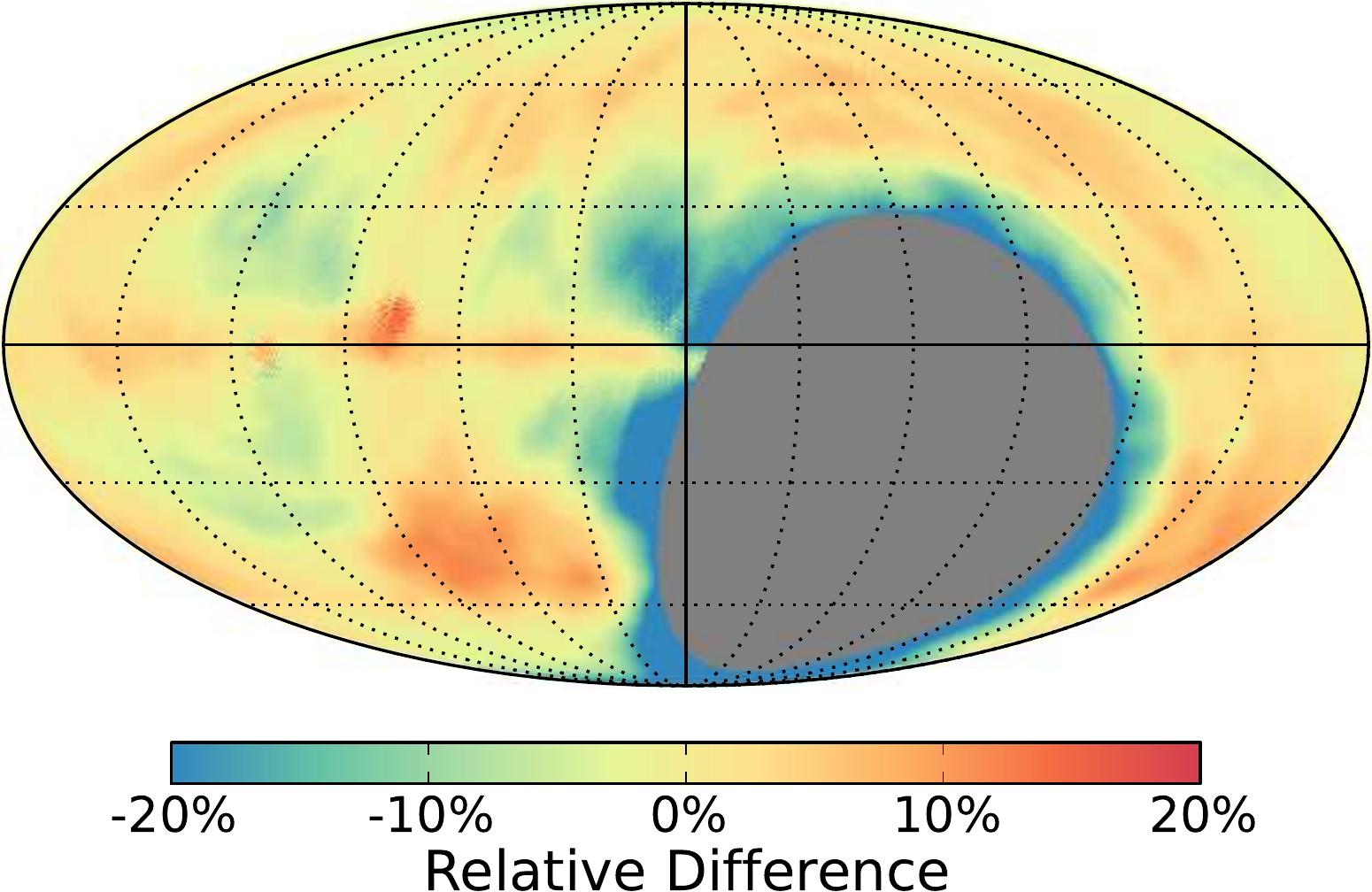}
	\caption{\label{fig:lfsmRatio}Comparison between our map at 74 MHz with a realisation of the LFSM.  The colour scale is similar to Figure \ref{fig:comp45} except that the colour scale saturates at $\pm$20\%.  Overall there is agreement between the data and the realisation at the 10\% level.  However, the region south of --20$^\circ$ declination appears to be systematically over predicted by the model relative to our map.}
\end{figure}

We have also examined the accuracy of the two dimensional structure in the model using the relative differencing procedure of Section \ref{sec:reldiff}.  Figure \ref{fig:lfsmRatio} shows the comparison at 74 MHz between a realisation of the LFSM and our map.  We find that the realisation is within $\sim$10\%, with most differences being within $\sim$20\%.  The most notable difference between the data and the realisation is in the area between --20 degrees declination and our horizon limit and the area near a Galactic latitude of --60$^\circ$.  The latter is likely related to the artefacts in our 74 MHz map noted in Section \ref{sec:reldiff}.

\section{Conclusions}
\label{sec:conc}
We have presented maps of the low frequency radio sky between 35 and 80 MHz using the first station of the Long Wavelength Array.  These maps were partly motivated by recent efforts to detect cosmological 21-cm features and represent the first wide band, self consistent survey of the sky created at these frequencies.  Based on various comparisons with existing surveys we conservatively estimate the calibration of these maps in the 15\% to 25\% range, both in terms of the resolved structure and the sky-averaged temperature.  We have also combined our maps with those from the literature to create an updated model of the sky between 10 and 408 MHz.  This new model, called the Low Frequency Sky Model, is a significant step towards a more accurate model of the low frequency radio emission.  In particular, the additional maps used in our model help to better constrain the contributions of free-free absorption from \ion{H}{2} regions that become increasingly more important towards lower frequencies and within the Galactic plane.

Although these maps provide a self consistent view of the sky, they do not span the full frequency range of LWA1.  In the future we plan to extend the maps in frequency, both lower and higher.  Toward the lower end we are limited by abundant RFI down to the ionospheric cutoff at about 10 MHz.  However, this RFI has been seen to have a strong diurnal variation and it may be possible to image the sky in an opportunistic fashion by collecting data exclusively at night.  The approach would also help in establishing if the structure seen at the Galactic poles is real or an artefact of low-level RFI.  At the high end of the frequency band it might be possible to create high fidelity images up to 85 MHz.  Here the challenge will be to overcome the drop in sensitivity of the system caused by the rolloff in the analog filters, possibly by doubling the data collection cadence.  Beyond collecting additional data, additional work is also needed in understanding the antenna impedance mis-match losses and the dipole gain pattern.  A detailed understanding of the mis-match losses, both for LWA1 and the LEDA system, is of particular value since these losses influence both the spectral shape of the measurements and the total power.  Studies to map the dipole gain pattern, particularly at the lower elevations, $\lesssim$20$^\circ$, would benefit from a higher resolution low frequency array, such as the LWA station in Owens Valley\footnote{\url{http://www.tauceti.caltech.edu/lwa/}}, which would have reduced confusion noise that would allow many more sources to be observed at all frequencies as they transit the sky.  The enhanced resolution of such an array would also be helpful in creating more detailed maps that can be used to help define a collection of low frequency flux calibrators, particularly at lower declinations.

Beyond these improvements to our survey there are also improvements that can be made to the sky model.  Models of the low frequency radio sky would benefit from modern all-sky surveys of the diffuse Galactic emission in the 100 to 200 MHz range that could be performed by instruments such as LOFAR and Murchison Widefield Array.  This would help constrain the models in the region needed for epoch of reionisation work and be generally useful for understanding the non-thermal emission from our own galaxy.  In addition to surveys at other frequencies, more sophisticated analysis techniques are needed in order to more accurately combine the survey data together and interpolate between surveys.  The enhanced principal component analysis of \citet{egsm} and the physically motivated decomposition of \citet{gmoss} are recent developments towards this goal.  The method of \citet{gmoss} is of particular interest since it directly relates the emission to the underlying physical processes.  More sophisticated analysis techniques may also be able to account for the two-dimensional uncertainties of the constituent surveys in order to provide a more robust model.  Furthermore, the combination of new data with more modern analysis techniques will also allow the possibility of systematic offsets in the calibration between various surveys to be explored.

\section*{Acknowledgements}
We thank Sanjay Bhatnagar for a productive conversation on wide field imaging and mosaicking.  We also thank the referee for numerous constructive comments.  Construction of the LWA has been supported by the Office of Naval Research under Contract N00014-07-C-0147. Support for operations and continuing development of the LWA1 is provided by the National Science Foundation under grants AST-1139963 and AST-1139974 of the University Radio Observatory programme.  This work has made use of LWA1 outrigger dipoles made available by the LEDA project, funded by NSF under grants AST-1106054, AST-1106059, AST-1106045, and AST-1105949

\bibliographystyle{mnras}
\bibliography{ms} 

\bsp	
\label{lastpage}
\end{document}